\documentclass[journal]{IEEEtran}
\usepackage{cite}
\usepackage{xcolor,soul,framed} %,caption
\colorlet{shadecolor}{yellow}
\usepackage{color,soul}
\usepackage[pdftex]{graphicx}
\graphicspath{{../pdf/}{../jpeg/}}
\DeclareGraphicsExtensions{.pdf,.jpeg,.png}
\usepackage[cmex10]{amsmath}
\usepackage{algorithmic}
\usepackage{mdwmath}
\usepackage{mdwtab}
\usepackage{eqparbox}
\usepackage{url}
\usepackage{array}
\usepackage{fixltx2e}
\usepackage{dblfloatfix}
\usepackage[mathscr]{euscript}
\usepackage{amsfonts}
\usepackage{amsmath}
\usepackage{caption}
\usepackage{commath}
\usepackage{upgreek}
\usepackage[long]{optidef}
\usepackage{bm}
\usepackage{algorithm}
\DeclareMathOperator{\vect}{vec}
\DeclareMathOperator*{\argmin}{arg\,min}
\usepackage{cleveref}
\usepackage{balance}
\usepackage{textgreek}
\usepackage{booktabs,chemformula}
\usepackage{array}
\usepackage{float}
\usepackage{amssymb}

\begin{document}

\title{Interference Management in UAV-assisted Integrated Access and Backhaul Cellular Networks}

\author{Abdurrahman~Fouda,~\IEEEmembership{Student Member,~IEEE,}
        Ahmed~S.~Ibrahim,~\IEEEmembership{Member,~IEEE,}
        {\.{I}}smail~G{\"{u}}ven{\c{c}}~\IEEEmembership{Senior~Member,~IEEE,}
        and~Monisha~Ghosh,~\IEEEmembership{Fellow,~IEEE}
    \thanks{This research is supported in part by U.S. National Science Foundation under grants No. CNS-1618692 and CNS-1618836.}
    \thanks{A. Fouda and A. S. Ibrahim are with the Department of Electrical and Computer Engineering, Florida International  University, Miami, FL, 33174 USA (email: {afoud004, aibrahim}@fiu.edu).} 
    \thanks{{\.{I}}.~G{\"{u}}ven{\c{c}} is with the Department of Electrical and Computer Engineering, North Carolina State University, Raleigh, NC, 27606 USA (email: iguvenc@ncsu.edu).}
    \thanks{M. Ghosh is with the Department of Computer Science, University of Chicago, Chicago, IL, 60637 USA (email: monisha@uchicago.edu).}}
%\markboth{Accepted for publication in {IEEE Access journal}}%

\maketitle
\vspace{-0.5in}
\begin{abstract}
An integrated access and backhaul (IAB) network architecture can enable flexible and fast deployment of next-generation cellular networks. However, mutual interference between access and backhaul links, small inter-site distance and spatial dynamics of user distribution pose major challenges in the practical deployment of IAB networks. To tackle these problems, we leverage the flying capabilities of unmanned aerial vehicles (UAVs) as hovering IAB-nodes and propose an interference management algorithm to maximize the overall sum rate of the IAB network. In particular, we jointly optimize the user and base station associations, the downlink power allocations for access and backhaul transmissions, and the spatial configurations of UAVs. We consider two spatial configuration modes of UAVs: distributed UAVs and drone antenna array (DAA), and show how they are intertwined with the spatial distribution of ground users. Our numerical results show that the proposed algorithm achieves an average of $2.9\times$ and $6.7\times$ gains in the received downlink signal-to-interference-plus-noise ratio (SINR) and overall network sum rate, respectively. Finally, the numerical results reveal that UAVs cannot only be used for coverage improvement but also for capacity boosting in IAB cellular networks.

\end{abstract}
\IEEEpeerreviewmaketitle
\section{Introduction}\label{sec_intro}
%***************PARAGRAPH MESSAGE START***********************
%IAB: a potential candidate for flexible deployment of next-generation cellular networks. 
%***************PARAGRAPH MESSAGE END*************************
\IEEEPARstart{I}{n} recent years, the concept of wireless backhauling has emerged as a potential solution to reduce the deployment cost of cellular networks~\cite{joint2018,mMIMOUDN}. In this regard, $\text{3}\text{rd}$ Generation Partnership Project (3GPP) has introduced the integrated access and backhaul (IAB) network architecture to allow for flexible deployment of next-generation cellular networks~\cite{3GPPIAB,3GPPIAB_att}. Generally, the IAB architecture implies tight interworking between access and backhaul links, where the IAB-donor (i.e., macro base station (MBS)) uses the same infrastructure and wireless channel resources to provide access and backhauling functionalities for cellular users and IAB-nodes (i.e., small bases stations (SBSs)), respectively~\cite{jointIBIAB,BWPart,zorziIAB}. Although IAB-based cellular networks are envisioned to meet the increase in user and traffic demands, the mutual interference between access and backhaul links and the limitations of backhaul capacity are among the main challenges to develop reliable communication links in IAB networks (see, e.g., \cite{jointIBIAB}). 

We consider the unmanned aerial vehicles (UAVs) as a promising candidate to tackle these challenges in the IAB-based cellular networks. In particular, we investigate the potential gains of leveraging the flying capabilities of UAVs as hovering IAB-nodes in UAV-assisted IAB networks. There have been several recent studies where utilizing UAVs is proposed as a cost-effective and easily-scalable solution that can achieve significant performance improvements in wireless networks~\cite{totUAV}. Moreover, unlike the basic idea of dense deployment of SBSs to get closer to edge users~\cite{5GTechs}, the use of UAVs allows for the network architecture to be reconfigured dynamically based on the coverage and capacity demands~\cite{UAV3D5G,cachingsky}. Having UAVs communicating towards MBSs over backhaul links and towards cellular users over access links naturally leads to creating a wirelessly backhauled network architecture~\cite{UAVIBIAB,BHUAV1BS,UAVBHMBSs}. Therefore, there has been great interest in studying the performance of UAVs on both the access and backhaul networks. 

%The access link performance gains of using UAVs have been studied extensively in the literature for public safety~\cite{PubSafUAV,UAVPS_GC}, device-to-device communications~\cite{UAVD2DCach,UAVD2D_GT,UAVD2DFD}, Internet of Things (IoT) applications~\cite{UAVIoT,IoTUAVSurvey,IoTUAVPlat}, smart cities~\cite{UAVITS_SC,UAVsSCs_IWCMC}, non-orthogonal multiple access (NOMA) in millimeter wave (mmWave) networks~\cite{UAVNOMAlimitFB, AngCSIUAVNOMA,UAVNOMADistCSI}, coverage~\cite{UAVCovg3DNet,UAVEnergyEff,3DCovQoS}, connectivity~\cite{OBIABUAV,UAVConnec,UAVConnec_PerAna,UAVConnec_Traj} and capacity maximization~\cite{UAVHetNet, EffUAV}. Furthermore, 3GPP has been investigating the integration of UAVs into existing cellular networks~\cite{3GPPUAV,E///UAVs3gpp}. In addition to the conventional spatial configuration of UAVs as distributed nodes, the array directivity gains of configuring a group of UAVs in a single drone antenna array (DAA) were presented in~\cite{DAASLL,UAVLoSMIMO,DAAJor}. On the backhaul network side, the limitation of backhaul transmission capacity in UAV-assisted networks was discussed in~\cite{BHUAV1BS,UAVBHMBSs}. However, these works have not considered tight interworking between access and backhaul links, along with the resulting inter-cell interference. 
The access link performance gains of using UAVs have been studied extensively in the literature for public safety~\cite{PubSafUAV,UAVPS_GC}, device-to-device communications~\cite{UAVD2DCach,UAVD2DFD}, Internet of Things (IoT) applications~\cite{UAVIoT,IoTUAVPlat}, smart cities~\cite{UAVITS_SC}, non-orthogonal multiple access (NOMA) in millimeter wave (mmWave) networks~\cite{AngCSIUAVNOMA,UAVNOMADistCSI}, coverage~\cite{UAVCovg3DNet,RayUAV,3DCovQoS}, connectivity~\cite{OBIABUAV,UAVConnec,UAVConnec_Traj} and capacity maximization~\cite{UAVHetNet, EffUAV}. Furthermore, 3GPP has been investigating the integration of UAVs into existing cellular networks~\cite{3GPPUAV,E///UAVs3gpp}. 
In addition to the conventional spatial configuration of UAVs as distributed nodes, the array directivity gains of configuring a group of UAVs in a single drone antenna array (DAA) were presented in~\cite{DAASLL,UAVLoSMIMO,DAAJor}. On the backhaul network side, the limitation of backhaul transmission capacity in UAV-assisted networks was discussed in~\cite{BHUAV1BS,UAVBHMBSs}. However, these works have not considered tight interworking between access and backhaul links, along with the resulting inter-cell interference. 

To the best of the authors' knowledge, none of the prior studies have considered the problem of optimizing the overall network performance in UAV-assisted IAB networks. In this paper, we propose an interference management algorithm to jointly optimize user-BS associations, downlink power allocations and the 3D deployment of UAVs in UAV-assisted IAB networks. In particular, we present two spatial configuration modes of UAVs; namely, distributed UAVs and DAA; based on the spatial dynamics of ground user distribution. Moreover, we consider in-band backhauling, as a natural candidate for tighter interworking between access and backhaul links. In the former configuration mode, we define the 3D deployment of UAVs. In the latter mode, we define the DAA design parameters in terms of array orientation, drone element separation and the 3D placement of array center. The problem is cast as a network sum rate maximization problem and decomposed into two subproblems due to the mutual dependence between the optimization variables. The first subproblem is solved using a two-stage fixed-point method to find user-BS associations and downlink power allocations for access and backhaul transmissions, given fixed UAV spatial configurations. The second subproblem is solved using particle swarm optimization (PSO) to define the spatial configurations of UAVs and update power allocations given fixed user-BS associations. 

Our numerical results show that the proposed algorithm achieves an average of $3.1\times$ and $6.7\times$ gains in received downlink signal-to-interference-plus-noise ratio (SINR) and overall network sum rate, respectively, compared to the baseline scenario, in which, UAVs are not used. We demonstrate that the use of UAVs in in-band IAB networks results in both coverage enhancement and capacity boosting. As for the DAA configuration, the numerical results also reveal that the achievable network performance gains are directly proportional to the number of drone elements in the DAA. In this regard, we show how the computational complexity of the proposed algorithm can be independent of the number of UAVs when they are configured as DAA. Finally we point out that in our earlier work~\cite{UAVIBIAB}, an exhaustive search-based approach was proposed to investigate the sum rate and coverage gains of using distributed UAVs in in-band IAB networks.

The rest of this paper is organized as follows. The system model of distributed UAVs configuration mode is presented in Section~\ref{sec_sysmod}. The problem formulation is described in Section~\ref{sec_OptProbform}. In Section~\ref{sec_optprobsol}, we discuss the proposed interference management algorithm. The DAA configuration mode is provided in Section~\ref{sec_DAA}. Section~\ref{sec_results} presents numerical evaluations of the proposed algorithm. Finally, concluding remarks are given in Section~\ref{sec_conc}.

\section{System Model of Distributed UAVs Spatial Configuration Mode}\label{sec_sysmod}
% =======
% FIG. 01
% =======
\begin{figure}
  \begin{center}
  \includegraphics[width=8.5cm,height=8.5cm,,keepaspectratio]{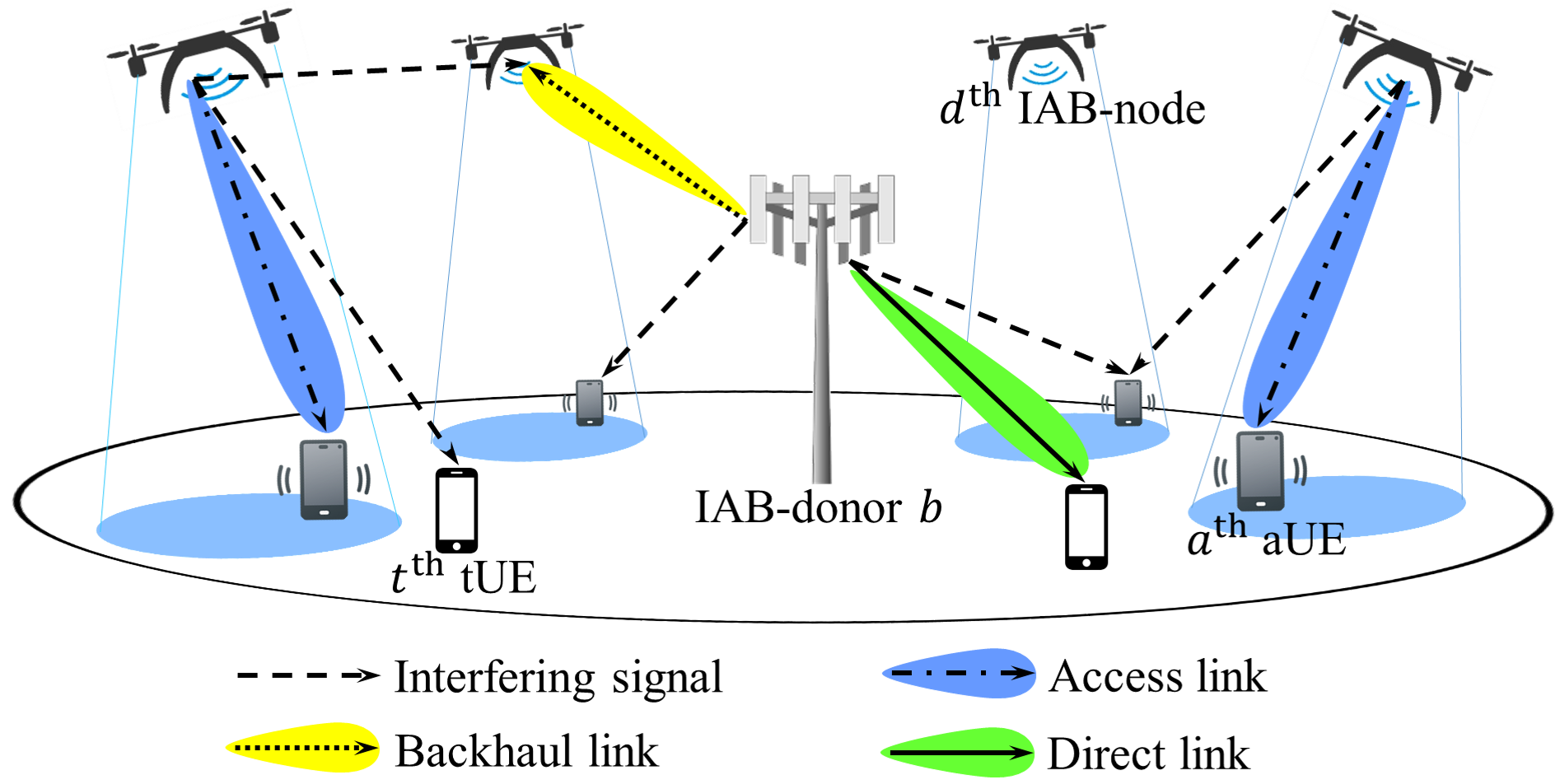}
  \caption{In-band IAB system architecture for next-generation cellular networks: UAVs can be users themselves or operate as drone IAB-nodes to serve other users.}\label{fig_sysmod}
  \vspace{-.25in}
  \end{center}
\end{figure}
%***************PARAGRAPH MESSAGE START***********************
%Next two paragraphs present thev2 definitions that we use in the proposed system model
%***************PARAGRAPH MESSAGE END*************************
Fig.~\ref{fig_sysmod} depicts the considered in-band UAV-assisted two-tier IAB network, in which the access and backhaul links fully overlap in spectrum resources~\cite{3GPPIAB}. The first tier represents the IAB-donor $b$ that supports $T$ terrestrial users (tUEs) with direct links and provides wireless backhauling functionality to $D$ UAVs. The second tier represents UAVs operating as drone IAB-nodes to support $A$ aerial users (aUEs) with access links. The downlink transmission denotes the data transmission from UAVs to aUEs and IAB-donor to tUEs and UAVs, respectively. The IAB-donor is equipped with $N$ element uniform linear array (ULA). UAVs and users are equipped with single receiving and transmitting antennas. Also, we assume spatial distribution of ground users into $D$ clusters. Let $\mathscr{D}=\left\{1,\dots,D\right\}$, $\mathscr{A}=\left\{1,\dots,A\right\}$ and $\mathscr{T}=\left\{1,\dots,T\right\}$ denote the sets of UAVs, aUEs and tUEs, respectively where, e.g., the cardinality of $\mathscr{D}$ is $\abs{\mathscr{D}}$ and is equal to $D$. The set of BSs is represented by $\mathscr{S}=\left\{1,\dots,S\right\}$ where $S=D+1$. Finally, the set of users is represented by $\mathscr{U}=\mathscr{A} \cup \mathscr{T}$ where $U=A+T$. 

The multiple-input-single-output (MISO) downlink channel ${\mathbf{h}_{\mathrm{b},t}}\in{\mathbb{C}^{1\times{N}}}$ between IAB-donor and $t^{\text{th}}$ tUE is introduced as~\cite[Ch.~7]{D.Tse}:
\begin{equation}\label{equ_hb>t}
    \mathbf{h}_{\mathrm{b},t} = \frac{1}{\sqrt{K}} \times \sum_{k=1}^{K} \frac{{g_{}}_{\mathrm{b}t,k}\mathbf{a}^*\left({\theta_{}}_{\mathrm{b}t,k}\right)}{1+\left(d_{\mathrm{b},t}\right)^\alpha}\text{,}
\end{equation}
where $K$, ${g_{}}_{\mathrm{b}t,k}$, ${\theta_{}}_{\mathrm{b}t,k}$, $d_{\mathrm{b},t}$ and $\alpha$ represent the number of propagation paths, complex channel gain of the $k^{\text{th}}$ path, angle-of-departure (AoD) of the $k^{\text{th}}$ path, 3D distance between IAB-donor and $t^{\text{th}}$ tUE and pathloss coefficient, respectively. ${g_{}}_{\mathrm{b}t,k}$ follows standard complex Gaussian distribution with $\mathcal{CN}\left(0,1\right)$ and ${\theta_{}}_{\mathrm{b}t,k}$ follows a uniform distribution with $\mathcal{U}\left[{\theta_{}}_{\mathrm{b},t}^{\text{LOS}}-\mathrm{ASD},{\theta_{}}_{\mathrm{b},t}^{\text{LOS}}+\mathrm{ASD}\right]$ where ${\theta_{}}_{\mathrm{b},t}^{\text{LOS}}$ is the line-of-sight (LOS) angle between IAB-donor and $t^{\text{th}}$ tUE, and $\mathrm{ASD}$ is the angular spread of departure and follows the same distribution as~\cite[Table~7.5-6]{SCM}. The transmit antenna array steering vector of the $k^{\text{th}}$ path and AoD ${\theta_{}}_{{\rm{b}}t,k}$ is given by:
\begin{equation}\label{equ_steering}
    \begin{split}
        &\mathbf{a}\left({\theta_{}}_{{\mathrm{b}}t,k}\right)=\\ &\frac{1}{\sqrt{N}} \left[1,e^{-j2\pi\frac{\Delta}{\lambda}\mathrm{sin}\left({\theta_{}}_{\mathrm{b}t,k}\right)},\dots, e^{-j2\pi\frac{\Delta}{\lambda}(N-1)\mathrm{sin}\left({\theta_{}}_{\mathrm{b}t,k}\right)}\right]^\top \!\!\!\text{,}
    \end{split}
\end{equation}
where $\Delta$ is the antenna element separation of the ULA and $\lambda$ is the carrier wavelength. Similarly, the backhaul channel between IAB-donor and $d^{\text{th}}$ drone is represented by ${\mathbf{h}_{\mathrm{b},d}}\in{\mathbb{C}^{{1}\times{N}}}$ and the access channel between $d^{\text{th}}$ drone and $a^{\text{th}}$ aUE is represented by ${h_{d,a}}\in{\mathbb{C}^{{1}\times{1}}}$.

\subsection{Backhaul Downlink Transmissions}\label{subsec_sysmod_BHDLTXions}
%***************PARAGRAPH MESSAGE START***********************
%Next paragraph explains the LZFBF precoder at the IAB-donor. 
%***************PARAGRAPH MESSAGE END*************************
We consider linear zero-forcing beamforming (LZFBF) for multi-user MISO transmissions at backhaul links, in which, the ZF precoder at IAB-donor is defined as ${\mathbf{V}_{\mathrm{b}}}\in{\mathbb{C}^{{N}\times(D+1)}}$, where $\mathbf{V}_\mathrm{b}=\mathbf{H}_\mathrm{b}^\dagger={\mathbf{H}}_{\mathrm{b}}^{\text{*}}{\left[{\mathbf{H}}_{\mathrm{b}}^{}{\mathbf{H}}_{\mathrm{b}}^{\text{*}}\right]}^{-1}$. The full rank channel matrix between IAB-donor, UAVs and $t^{\text{th}}$ tUE scheduled at $f^{\text{th}}$ subcarrier and $l^{\text{th}}$ time slot is given by ${\mathbf{H}_\mathrm{b}(f,l)}\in{\mathbb{C}^{{(D+1)}\times{N}}}$ where ${\mathbf{H}_{\mathrm{b}}(f,l)}=\left[\mathbf{h}_{\mathrm{b},1}(f,l),\dots,\mathbf{h}_{\mathrm{b},D}(f,l),\mathbf{h}_{\mathrm{b},t}(f,l)\right] $. For simplicity of presentation, we omit references to $(f,l)$ indices in the rest of this paper. The precoding vector between IAB-donor and $i^{\text{th}}$ reception point is normalized using equal transmit power (ETP) normalization due to its higher sum rate gains~\cite{ETP}, and is given by $\mathbf{v}_{\mathrm{b},i}=\left[\mathbf{V}_\mathrm{b}\right]_i^{}/ \norm{\left[\mathbf{V}_\mathrm{b}\right]_{i}}$, where $\left[\mathbf{V}_\mathrm{b}\right]_{i}$ is the $i^{\text{th}}$ column of $\mathbf{V}_\mathrm{b}$. 

The received signal at $d^{\text{th}}$ drone from IAB-donor (see Fig.~\ref{fig_sysmod}) can be modeled as: 
\begin{equation}\label{equ_b>d}
    \begin{split}
        {y_{}}_{\mathrm{b},d}&=\underbrace{\sqrt{{p_{}}_{\mathrm{b},d}}{\mathbf{h}_{}}_{\mathrm{b},d}{\mathbf{v}_{}}_{\mathrm{b},d}{x_{}}_{\mathrm{b},d}}_{\text{transmitted signal}}\\&+\underbrace{\sum_{{i}\in{{\overline{\mathscr{A}}_{}}_d}}\sqrt{{p_{}}_{d,i}}{h_{}}_{d,d}{x_{}}_{d,i}}_{\text{self-interference}}+\underbrace{\sum_{{i}\in{{\overline{\mathscr{A}}_{}}_\mathrm{b}}}\sqrt{{p_{}}_{\mathrm{b},i}}{h_{}}_{\mathrm{b},d}{\mathbf{v}_{}}_{\mathrm{b},i}{x_{}}_{\mathrm{b},i}}_{\text{inter-stream interference}}\\&+\underbrace{\sum_{{j}\in{\mathscr{D}\backslash d}}\sum_{{i}\in{{\overline{\mathscr{A}}_{}}_j}} \sqrt{{p_{}}_{j,i}}{h_{}}_{j,d}{x_{}}_{j,i}}_{\text{inter-tier interference}}+{n_{}}_d\text{,}
    \end{split}
\end{equation}
where ${p_{}}_{\mathrm{b},d}$, ${\mathbf{v}_{}}_{\mathrm{b},d}$ and $x_{\mathrm{b},d}$ represent the backhaul downlink power allocation, precoding vector and transmitted data symbol. ${\overline{\mathscr{A}}_{}}_d$ and ${\overline{\mathscr{A}}_{}}_j$ denote the sets of interfering aUEs that are associated with $d^{\text{th}}$ and $j^{\text{th}}$ UAVs, respectively where ${j}\neq{d}$. The second, third and fourth terms in~(\ref{equ_b>d}) represent the self-interference between access and backhaul, inter-stream interference and inter-tier interference on backhaul transmissions of $d^{\text{th}}$ drone. $n_d\sim\mathcal{CN}(0,\,\sigma^{2})$ denotes the received zero-mean complex Gaussian noise with variance {$\sigma^{2}$} at $d^{\text{th}}$ UAV. Each UAV is a full-duplex capable drone IAB-node, which can be integrated into in-band IAB scenarios without self-interference constraints. We assume perfect channel state information (CSI) knowledge at IAB-donor. Further, LZFBF is used to suppress the inter-stream interference between $\left(D+1\right)$ independent spatial streams of backhaul and direct links \cite{ZFDetc,ZFBFGold}. Hence, the second and third terms can be omitted from~(\ref{equ_b>d}) and the received SINR at $d^{\text{th}}$ drone can be calculated as:
\begin{equation}
    \label{equ_SINR_b>d}
        {\upgamma_{}}_{\mathrm{b},d}=\frac{{p_{}}_{\mathrm{b},d}\abs{{\mathbf{h}_{}}_{\mathrm{b},d}{\mathbf{v}_{}}_{\mathrm{b},d}}^{2}}{\displaystyle\sum_{j\in\mathscr{D}\backslash d}\abs{{h_{}}_{j,d}}^{2}\sum_{i\in{\mathscr{A}_{}}_j}p_{j,i}+\sigma^{2}}\text{.}
\end{equation}

\subsection{Access Downlink Transmissions}\label{subsec_sysmod_AccDLTXions}
%***************PARAGRAPH MESSAGE START***********************
%Next paragraphs present the received downlink signals in the proposed system model
%***************PARAGRAPH MESSAGE END*************************
Similarly, the received downlink signal at $t^{\text{th}}$ tUE from IAB-donor is given by:  
\begin{equation}\label{equ_b>tUE}
    {y_{}}_{\mathrm{b},t}=\underbrace{\sqrt{{p_{}}_{\mathrm{b},t}}{\mathbf{h}_{}}_{\mathrm{b},t}{\mathbf{v}_{}}_{\mathrm{b},t}{x_{}}_{\mathrm{b},t}}_{\text{transmitted signal}}+\underbrace{\sum_{j \in \mathscr{D}}\sum_{i\in \overline{{\mathscr{A}}_{}}_j}\sqrt{{p_{}}_{j,i}}{h_{}}_{j,t}{x_{}}_{j,i}}_{\text{inter-tier interference}}+{n_{}}_t\text{,}
\end{equation}
where $\overline{\mathscr{A}}_j$ is the set of associated aUEs with $j^{\text{th}}$ drone and are scheduled on same spectrum and time resources as $t^{\text{th}}$ tUE. The second term in~(\ref{equ_b>tUE}) represents the inter-tier interference on the access transmissions of $t^{\text{th}}$ tUE. The received SINR at $t^{\text{th}}$ tUE can be expressed by:
\begin{equation}
        \label{equ_SINR_b>t}
        {\upgamma_{}}_{\mathrm{b},t} =\frac{{p_{}}_{\mathrm{b},t}\abs{{\mathbf{h}_{}}_{\mathrm{b},t}{\mathbf{v}_{}}_{\mathrm{b},t}}^{2}}{\displaystyle\sum_{j\in\mathscr{D}}\abs{{h_{}}_{j,t}}^{2}\sum_{i\in{\overline{\mathscr{A}}_{}}_j}{p_{}}_{j,i}+\sigma^{2}}\text{.}
\end{equation} 

Finally, the received downlink signal at $a^{\text{th}}$ aUE from $d^{\text{th}}$ drone is given by:
\begin{equation}\label{equ_d>aUE}
    \begin{split}
        {y_{}}_{d,a} &=\underbrace{\sqrt{{p_{}}_{d,a}}{h_{}}_{d,a}{x_{}}_{d,a}}_{\text{transmitted signal}}+\underbrace{\sum_{j\in \mathscr{D}\backslash d} \sum_{i\in\overline{\mathscr{A}_{}}_j} \sqrt{{p_{}}_{j,i}}{h_{}}_{j,a}{x_{}}_{j,i}}_{\text{intra-tier interference}}\\&+\underbrace{\sum_{k\in\mathscr{D}\cup\overline{\mathscr{T}}}\sqrt{{p_{}}_{\mathrm{b},k}}{\mathbf{h}_{}}_{\mathrm{b},a}{\mathbf{v}_{}}_{\mathrm{b},k}{x_{}}_{\mathrm{b},k}}_{\text{inter-tier interference}}+n_a\text{,}
    \end{split}
\end{equation}
where $\mathscr{D}\cup\overline{\mathscr{T}}$ is the set of UAVs and tUEs scheduled on the same spectrum and time resources as $a^{\text{th}}$ aUE. The second and third terms in~(\ref{equ_d>aUE}) represent the intra-tier interference and inter-tier interference of the IAB-donor transmissions on the access transmissions of $a^{\text{th}}$ aUE. The received SINR at $a^{\text{th}}$ aUE is represented by: 
\begin{equation}
    \label{equ_SINR_d>a}
    \begin{split}
        {\upgamma_{}}_{d,a}=\\&\frac{{p_{}}_{d,a}\abs{{h_{}}_{d,a}}^{2}}{\displaystyle\sum_{j\in\mathscr{D}\backslash d}\abs{{h_{}}_{j,a}}^{2}\sum_{i\in{\overline{\mathscr{A}}_{}}_j}{{{p_{}}_{j,i}}}+ \sum_{k\in\mathscr{D}\cup\overline{\mathscr{T}}} {{p_{}}_{\mathrm{b},k}}\abs{\mathbf{h}_{\mathrm{b},a}{\mathbf{v}_{}}_{\mathrm{b},k}}^{2}+\sigma^{2}}\text{.}
    \end{split}
\end{equation}
\section{Problem Formulation}\label{sec_OptProbform}
%***************PARAGRAPH MESSAGE START***********************
%problem formulation
%***************PARAGRAPH MESSAGE END*************************
In this section, we formulate the joint optimization of user-BS associations, downlink power allocations and the 3D deployment of UAVs. To this end, The problem is cast as a network sum rate maximization problem subject to a received SINR threshold at each reception point and taking into account the transmission power constraint at each BS. The network sum rate maximization problem can be written as:
\begin{equation}\label{equ_PA}
    \mathcal{P:} \max_{\mathbf{C}, \mathbf{w}, \mathbf{p}, {\mathbf{p}_{}}_{\mathrm{BH}}} \quad \mathbf{1}_A^\top \log_2\left(1+{\boldsymbol\upgamma_{}}_\mathscr{A}\right)+\mathbf{1}_T^\top \log_2\left(1+{\boldsymbol\upgamma_{}}_\mathscr{T}\right)\text{,}\\
\end{equation}
\begin{subequations}
    \begin{align} 
        \label{equ_PA_sub1}
            \mathrm{subject \, to} \quad & {\boldsymbol{\upgamma}_{}}_\mathrm{\mathscr{U}}\ge{{\epsilon}_{}}_{\mathrm{u}}\text{,\,} {\boldsymbol\upgamma_{}}_\mathscr{D}\ge{\epsilon_{}}_{\mathrm{d}}\text{,}\\ 
        \label{equ_PA_sub2}
            &c_d\in\left[c_d^{(\mathrm{min})},c_d^{(\mathrm{max})}\right]\text{,\,}\forall \, c \in\ \left\{x,y,z\right\}\text{,}\\
        \label{equ_PA_sub3}
            &\mathbf{m}\le{\mathbf{p}_{}}_{\mathscr{S}}^{(\mathrm{max})}\text{,}
    \end{align}
\end{subequations}
where $\mathbf{1}_{A}$ denotes the $A$-dimensional all-ones vector, ${\boldsymbol\upgamma_{}}_\mathscr{A}=\left({\upgamma_{}}_{a}:a\in\mathscr{A}\right)$ and ${\boldsymbol\upgamma_{}}_\mathscr{T}=\left({\upgamma_{}}_{t}:t\in\mathscr{T}\right)$ denote the vectors of received downlink SINR at aUEs and tUEs, respectively. ${\mathbf{C}}\in{\mathbb{R}^{3 \times D}}$ denotes the 3D locations of UAVs with ${\mathbf{c}_{}}_d=\left[x_d,y_d,z_d\right]^\top$. The user-BS association vector is given by $\mathbf{w}\in\mathbb{R}^{1 \times U}$ where $\mathbf{w}=\left({w_{}}_{s,u}:u\in\mathscr{U}\right)$ contains the indices of serving BS of each user with value ${w_{}}_{s,u}:=s,\,{s}\in{\mathscr{S}}$. The user power allocation vector is given by ${\mathbf{p}}\in{\mathbb{R}^{1 \times U}}$, where $\mathbf{p}=\left({p_{}}_{s,u}:u\in\mathscr{U}\right)$ with ${p_{}}_{s,u}$ being the power allocated by $s^{\text{th}}$ BS for downlink transmissions of $u^{\text{th}}$ user based on association vector $\mathbf{w}$. Similarly, the UAV backhaul link power allocation vector is given by ${\mathbf{p}_{}}_{\mathrm{BH}}\in\mathbb{R}^{1\times D}$, where ${\mathbf{p}_{}}_{\mathrm{BH}}=\left({p_{}}_{b,d}:d\in\mathscr{D}\right)$.

Essentially, a low-quality backhaul link will bottleneck the access link. In this paper, we implement such dependency between the backhaul and access links in a binary fashion, as shown in the inequality constraint~(\ref{equ_PA_sub1}). In that, there will be no access transmissions if the received SINR levels at backhaul links are below a predefined threshold. It is worth noting that this binary dependency resembles selective decode-and-forward (DF) relaying mode, in which, the relay only forwards the signal if the received SINR exceeds a given threshold~\cite{TseRelay}. Finally, the boundaries of the feasible set of solutions are given by~(\ref{equ_PA_sub2}) and (\ref{equ_PA_sub3}). In that, the total power allocation vector of BSs is represented by ${\mathbf{m}}\in{\mathbb{R}^{1 \times S}}$ with $\mathbf{m}=\left(m_s:{s}\in{\mathscr{S}},m_s=\mathbf{1}^{\top}\mathbf{p}_s\right)$ where ${\mathbf{p}_{}}_{s}=\left({p_{}}_{s,i}:{i}\in{\mathscr{A}_{s}}\right)$ and $\mathscr{A}_s$ denote the power allocation vector and the total number of attached users to $s^{\text{th}}$ BS, respectively. The transmission power constraints of BSs are given by ${\mathbf{p}_{}}_{\mathscr{S}}^{(\mathrm{max})}=\left(p_s^{(\mathrm{max})}:{s}\in{\mathscr{S}}\right)$.

%***************PARAGRAPH MESSAGE START***********************
%Why do we decompose the optimization problem in (10) into two subproblems. %Summary of the solutions of PA and PB
%***************PARAGRAPH MESSAGE END*************************
According to the channel model in (\ref{equ_hb>t}), logarithmic objective function in (\ref{equ_PA}) and SINR constraints in (\ref{equ_PA_sub1}), the problem is considered as NP-hard mixed-integer nonlinear program (NP-MINLP)~\cite{FPIDLMaxMin}. Moreover, the problem cannot be considered as a single optimization problem due to the mutual dependence between the optimization variables. Essentially, increasing the downlink power allocations increases the received levels of signal power at cellular users and UAVs. However, given~(\ref{equ_SINR_b>d}),~(\ref{equ_SINR_b>t}) and~(\ref{equ_SINR_d>a}), the received levels of inter-tier and intra-tier interference increase as we increase the downlink power allocations. We also note that each suboptimal set of 3D locations of UAVs leads to different suboptimal sets of user-BS associations and power allocations. Hence, we solve the master optimization problem~(\ref{equ_PA}) to find the near-optimal set of power allocations, user-BS associations and 3D deployment of UAVs to maximize the received overall network downlink throughput while keeping the minimum levels of interference at access and backhaul links. 

To this end, and to make the optimization problem tractable, we decompose the mater problem in~(\ref{equ_PA}) into two subproblems denoted by $\mathcal{{PA}}$ and $\mathcal{PB}$. In $\mathcal{{PA}}$, we jointly optimize the user-BS associations and power allocations for access and backhaul downlink transmissions given fixed UAV spatial configurations. $\mathcal{{PA}}$ can be written as follows:

\begin{equation}\label{equ_PAPA}
    \begin{split}
        \mathcal{PA:} &\min_{\mathbf{w}, \mathbf{p}, {\mathbf{p}_{}}_{\mathrm{BH}}} \quad \mathbf{1}_U^\top \mathbf{p}+\mathbf{1}_D^\top {\mathbf{p}_{}}_{\mathrm{BH}}\text{,}\\&\mathrm{subject\,to}~\text{(\ref{equ_PA_sub1}) and (\ref{equ_PA_sub3})}\text{.}
    \end{split}
\end{equation}

%\begin{equation}\label{equ_PAPA}
%    \begin{split}
%        \mathcal{PA:} &\max_{\mathbf{w}, \mathbf{p}, {\mathbf{p}_{}}_{\mathrm{BH}}} \quad \mathbf{1}_A^\top \log_2\left(1+{\boldsymbol\upgamma_{}}_\mathscr{A}\right)+\mathbf{1}_T^\top \log_2\left(1+{\boldsymbol\upgamma_{}}_\mathscr{T}\right)\text{,}\\&\mathrm{subject\,to}~\text{(\ref{equ_PA_sub1}) and (\ref{equ_PA_sub3})}\text{.}
%    \end{split}
%\end{equation}

In $\mathcal{PB}$, we define the 3D hovering locations of UAVs and update downlink power allocations accordingly, given fixed user-BS associations. The subproblem $\mathcal{PB}$ is given by:
\begin{equation}\label{equ_PAPB}
    \begin{split}
        \mathcal{PB:} &\max_{\mathbf{C}, \mathbf{p}, {\mathbf{p}_{}}_{\mathrm{BH}}} \quad \mathbf{1}_A^\top \log_2\left(1+{\boldsymbol\upgamma_{}}_\mathscr{A}\right)+\mathbf{1}_T^\top \log_2\left(1+{\boldsymbol\upgamma_{}}_\mathscr{T}\right)\text{,}\\&\mathrm{subject\,to}~\text{(\ref{equ_PA_sub1}) - (\ref{equ_PA_sub3})}\text{.}
    \end{split}
\end{equation}

\section{Hybrid Fixed-Point Iteration and Particle Swarm Approach}\label{sec_optprobsol}
First we exploit fixed-point method and PSO to solve $\mathcal{{PA}}$ and $\mathcal{{PB}}$, respectively. An iterative algorithm is then presented to jointly optimize user-BS associations, power allocations and the 3D locations of UAVs by exploiting $\mathcal{{PA}}$ and $\mathcal{PB}$. The proposed algorithm converges to a near-optimal feasible set of solutions after a finite number of iterations. The optimization variables are updated every update time instant the network reaches a predefined user-drop rate, or when the quality of service (QoS) of a certain group of users decreases below a predetermined level. 

\subsection{Fixed-point Iteration Method for $\mathcal{PA}$}\label{subsec_optprobsol_fpi}
%***************PARAGRAPH MESSAGE START***********************
%Next paragraphs present the proposed fixed-point method. The algorithm is summarized in Algorithm 1 at the end of this section.
%***************PARAGRAPH MESSAGE END*************************
Let us first consider uniform random initialization for user-BS associations, where $\mathbf{w}(0)\sim\mathcal{U}\left[1,D\right]$. Similarly, the UAV 3D location matrix is initialized with uniformly distributed random locations  between ${c_{}}_{d}^{(\mathrm{min})}$ and $,{c_{}}_{d}^{(\mathrm{max})}$, where $\mathbf{C}(0)\sim\mathcal{U}\left[ {c_{}}_{d}^{(\mathrm{min})},{c_{}}_{d}^{(\mathrm{max})}\right]$. The downlink access and backhaul power allocations are also initialized with equal allocations based on the number of associated users with each BS, where ${p}_{s,u}:={{p_{}}_{s}^{(\mathrm{max})}}/\mathscr{A}_{s}$ and ${p_{}}_{\mathrm{b},d}:={p_{}}_{\mathrm{b}}^{(\mathrm{max})}/{\mathscr{A}_{}}_{\mathrm{b}}$. Now, let $t_{s,u}$ be the required power to receive unity SINR when $u^{\text{th}}$ user is associated with $s^{\text{th}}$ BS. In other words, given~(\ref{equ_SINR_b>t}) it can be calculated at $t^{\text{th}}$ tUE as:
\begin{equation}\label{ex_untsinr}
{t_{}}_{\mathrm{b},t}=\frac{\sum_{j\in\mathscr{D}}\abs{{h_{}}_{j,t}}^{2}\sum_{i\in{\overline{\mathscr{A}}_{}}_j}{p_{}}_{j,i}+\sigma^{2}}{{\abs{{\mathbf{h}_{}}_{\mathrm{b},t}{\mathbf{v}_{}}_{\mathrm{b},t}}^{2}}}\text{.}    
\end{equation}

Hence, the matrix of required power allocations to have a unity SINR at all users can be written as ${\mathbf{T}_{\mathrm{u}}}\in{\mathbb{R}^{S \times U}}$, where $t_{s,u}$ denotes the value of element $\mathbf{T}_{\mathrm{u}}\lbrack s,u\rbrack$. In other words, ${\mathbf{T}_{\mathrm{u}}}$ calculates the required power allocation at $u^{\text{th}}$ user to receive a unity SINR when it is associated with $s^{\text{th}}$ BS $\forall s\in{\mathscr{S}}$. The optimum power allocation at each user is defined as the minimum power allocation among all BSs. Hence, the user-power allocation vector of $(i+1)^{\text{th}}$ iteration can be updated as: 
\begin{equation}\label{equ_FPI_user-PwrAlloc}
    \mathbf{p}(i+1)\gets{\displaystyle\min_{s\in\mathscr{S}}t_{s,u}(i)}, u\in\mathscr{U}\text{,}
\end{equation}
where $\mathbf{p}$ is a vector of column-minima of $\mathbf{T}_{\mathrm{u}}$. The corresponding user-BS association can be given accordingly by:
\begin{equation}\label{equ_FPI_user-BSAss}
    \mathbf{v}(i+1)\gets{\displaystyle \argmin_{s\in\mathscr{S}}t_{s,u}(i)}, u\in\mathscr{U}\text{.}
\end{equation}
Similarly, the required backhaul power allocations to receive unity SINR at UAVs is denoted by ${\mathbf{t}_{\mathrm{BH}}(i+1)}\in{\mathbb{R}^{1 \times D}}$ and is computed based on the association vector $\mathbf{v}(i+1)$.  

Now, for $t^{\mathrm{th}}$ tUE to receive a minimum SINR of $\epsilon_{\mathrm{u}}$, the user power allocation in~(\ref{ex_untsinr}) can be updated as ${t_{}}_{\mathrm{b},t}\gets{{\epsilon_{}}_{\mathrm{u}}}{t_{}}_{\mathrm{b},t}$. In other words, if a power allocation of ${t_{}}_{\mathrm{b},t}$ gives an $\mathrm{SINR}=1$, then a power allocation of ${{\epsilon_{}}_{\mathrm{u}}}{t_{}}_{\mathrm{b},t}$ gives an $\mathrm{SINR}={\epsilon_{}}_{\mathrm{u}}$. Given that $\mathbf{p}(i+1)\in{\mathbb{R}^{1 \times U}}$ in (14) denotes the optimum user-power allocation vector of $(i+1)^{\text{th}}$ iteration to reach a unity SINR at each user, it can be updated as follows: 
\begin{equation}\label{equ_FPI_pwrupd}
    \mathbf{p}(i+1)\gets{{\epsilon_{}}_{\mathrm{u}}\mathbf{p}(i+1)}\text{,}
\end{equation}
in order to receive a minimum SINR of ${\epsilon_{}}_{\mathrm{u}}$ at all users. Similarly, given that ${\mathbf{t}_{\mathrm{BH}}(i+1)}\in{\mathbb{R}^{1 \times D}}$ is the optimum backhaul power allocations of $(i+1)^{\text{th}}$ iteration to receive unity SINR at UAVs, the backhaul power allocation vector ${\mathbf{p}_{}}_{\mathrm{BH}}\in\mathbb{R}^{1\times D}$ can be updated as follows:
\begin{equation}\label{equ_FPI_pwrupd}
    {\mathbf{p}_{}}_{\mathrm{BH}}(i+1)\gets{{\epsilon_{}}_{\mathrm{d}}{\mathbf{t}_{}}_{\mathrm{BH}}(i+1)}\text{,}
\end{equation}
in order to receive a minimum SINR of ${\epsilon_{}}_{\mathrm{d}}$ at all UAVs. For simple notations, we omit references to index $i$ throughout the rest of this section.

Next, we adjust the updated user and backhaul power allocations based on the total power allocations of each BS to satisfy the inequality constraint in (\ref{equ_PA_sub3}). First, the user power allocations in (\ref{equ_FPI_pwrupd}) are adjusted using the following fixed-point equation:  
\begin{equation}\label{equ_FPI_userupd_fixed1}
    \mathbf{p}=\min\left\{\epsilon_{\mathrm{u}}\mathbf{p},\,\mathbf{p}_{\mathscr{S}}^{(\mathrm{lim})}\right\}\text{,}
\end{equation}
where ${\mathbf{p}_{}}_{\mathscr{S}}^{(\mathrm{lim})}$ is the vector of maximum allowed transmission power of BSs and is given by ${\mathbf{p}_{}}_{\mathscr{S}}^{(\mathrm{lim})}={\mathbf{p}_{}}_{\mathscr{S}}^{(\mathrm{max})}\oslash\mathbf{A}_{\mathscr{S}}$. $\mathbf{A}_{\mathscr{S}}$ contains the number of associated users to each BS where $\mathbf{A}_{\mathscr{S}}=(\mathscr{A}_{s}:s\in\mathscr{S})$ and $\oslash$ denotes the Hadamard division. Second, the proposed fixed-point algorithm follows a two-stage procedure to adjust the backhaul power allocations in (\ref{equ_FPI_pwrupd}). Let us consider the maximum allowed backhaul power allocation as: 
\begin{equation}\label{equ_FPI_BHmax}
    \text{\textzeta}=\frac{{p_{}}_{\mathrm{b}}^{(\mathrm{max})}-\displaystyle\sum_{i\in{\mathscr{A}_{}}_{\mathrm{b}}}{p_{}}_{\mathrm{b},i}}{D}\text{,}
\end{equation}
where ${p_{}}_{\mathrm{b}}^{(\mathrm{max})}$ and ${\mathscr{A}_{}}_{\mathrm{b}}$ are the transmission power constraint of IAB-donor and the set of associated tUEs with IAB-donor, respectively. Hence, the backhaul power allocations can be adjusted using following fixed-point equation:
\begin{equation}\label{equ_FPI_BHupd_fixed2}
    \mathbf{p_{}}_{\mathrm{BH}}=\min\left\{{\epsilon_{}}_{d}{\mathbf{p}_{}}_{\mathrm{BH}},\,\text{\textzeta}\right\}\text{,}
\end{equation}
if ${{p_{}}_{b,d}}{\,}\forall{\,}{d\in\mathscr{D}}$ exceeds \textzeta. Otherwise, the backhaul power allocations are adjusted using the same procedure in (\ref{equ_FPI_userupd_fixed1}). 

The two-stage backhaul power allocation update procedure exploits the transmission power upper bound of the IAB-donor and assures that the inequality constraints of backhaul transmissions in~(\ref{equ_PA_sub1}) are satisfied, which is critical for UAV-assisted IAB scenarios. It also assures a global convergence to optimum power allocations and user-BS associations after finite number of iterations. Following the same argument in~\cite[Theorem~3]{FPIConv}, the proposed fixed-point method converges to a global optimal solution at a geometric rate with $\norm{\mathbf{p}_{c}(i)-\mathbf{p}_{c}^{*}}_{\infty}<{Ck^{i}}$, where $\norm{~.~}_{\infty}$ is the $\ell_{\infty}$-norm, $\mathbf{p}_{c}(i)$ is the combined user and backhaul power allocation vector generated by Algorithm~\ref{Algo_1} at iteration $i$ with $\mathbf{p}_{c}(i)=\left[\mathbf{p}(i),\mathbf{p}_{\mathrm{BH}}(i)\right]$, $\mathbf{p}_{c}^{*}$ is the optimal power allocations of $\mathcal{PA}$, and $C>0$ and $0<k<1$ are constants that depend on the problem settings (i.e., channel realizations, user locations and number of users and BSs). The fixed-point algorithm is summarized in Algorithm \ref{Algo_1}. 

\begin{algorithm}
\caption{Defines power allocations and user-BS associations given fixed UAV 3D locations. }\label{Algo_1}
\begin{algorithmic}[1]  
	\STATE\textbf{Inputs:} user positions, $\mathbf{C}$, $U$, $D$, $\mathbf{p}_{\mathscr{S}}^{(\mathrm{max})}$, maximum iterations $I_m$, convergence coefficient $j=1$, iteration number $i=1$
	\STATE\textbf{Initialization:} \\
	$\mathbf{w}(0)\sim\mathcal{U}\left[1,D\right]$, $\mathbf{p}(0)\gets {p}_{s,u}:={{p_{}}_{s}^{(\mathrm{max})}}/\mathscr{A}_{s}$, ${\mathbf{p}_{}}_{\mathrm{BH}}(0)\gets {p_{}}_{\mathrm{b},d}:={p_{}}_{\mathrm{b}}^{(\mathrm{max})}/{\mathscr{A}_{}}_{\mathrm{b}}$, $\mathbf{C}(0)\sim\mathcal{U}\left[ {c_{}}_{d}^{(\mathrm{min})},{c_{}}_{d}^{(\mathrm{max})}\right]$ 
	\STATE $\mathbf{p}(i)=\mathbf{p}(0)$, ${\mathbf{p}_{}}_{\mathrm{BH}}(i)={\mathbf{p}_{}}_{\mathrm{BH}}(0)$, $\mathbf{w}(i)=\mathbf{w}(0)$
	\WHILE{$j,i \leq I_m$}
	\STATE Compute $\mathbf{T}_{\mathrm{u}}(i)$
	\STATE $\mathbf{p}(i+1)\gets{\displaystyle\min_{s\in\mathscr{S}}t_{s,u}(i)}$
	\STATE $\mathbf{v}(i+1)\gets{\displaystyle \argmin_{s\in\mathscr{S}}t_{s,u}(i)}$
	\STATE Compute $\mathbf{t}_{\mathrm{BH}}(i)$
	\STATE $\mathbf{p}(i+1)\gets{{\epsilon_{}}_{\mathrm{u}}\mathbf{p}(i+1)},\,{\mathbf{p}_{}}_{\mathrm{BH}}(i+1)\gets{{\epsilon_{}}_{d}{\mathbf{t}_{}}_{\mathrm{BH}}(i)}$
		\IF {$\mathbf{m}\succ{\mathbf{p}_{}}_{\mathscr{S}}^{(\mathrm{max})}$}
		\STATE $\mathbf{p}(i+1)=\displaystyle\min\left\{{\epsilon_{}}_{\mathrm{u}}\mathbf{p}(i+1),{\mathbf{p}_{}}_{\mathscr{S}}^{(\mathrm{lim})}(i)\right\}$ 
		    \IF {${\mathbf{p}_{}}_{\mathrm{BH}}(i+1)> \text{\textzeta}(i)$}
		    \STATE ${\mathbf{p}_{}}_{\mathrm{BH}}(i+1)=\min\left\{{\epsilon_{}}_{\mathrm{d}}{\mathbf{p}_{}}_{\mathrm{BH}}(i+1),\text{\textzeta}(i)\right\}$
		    \ELSE
		    \STATE ${\mathbf{p}_{}}_{\mathrm{BH}}(i+1)=\min\left\{{\epsilon_{}}_{\mathrm{d}}{\mathbf{p}_{}}_{\mathrm{BH}}(i+1),{p_{}}_{\mathrm{b}}^{(\mathrm{lim})}\right\}$
		    \ENDIF
		\ENDIF
		\STATE\textbf{Convergence check:}
		\IF {$\norm{\mathbf{p}(i)-\mathbf{p}(i+1)}_{\infty}\leq{\epsilon_{1}}$,\\ $\norm{{\mathbf{p}_{}}_{\mathrm{BH}}(i)-{\mathbf{p}_{}}_{\mathrm{BH}}(i+1)}_{\infty}\leq{\epsilon_{2}}$,\\ $\norm{\mathbf{w}(i)-\mathbf{w}(i+1)}_{\infty}\leq{\epsilon_{3}}$ and (\ref{equ_PA_sub1}) for some $\epsilon_{i}>0$}
		\STATE $j=0$
		\ENDIF
		\STATE $\mathbf{p}(i)\gets\mathbf{p}(i+1)$, ${\mathbf{p}_{}}_{\mathrm{BH}}(i)\gets{\mathbf{p}_{}}_{\mathrm{BH}}(i+1)$, $i\gets{i+1}$
	\ENDWHILE
	\RETURN $\mathbf{w}(I_{f})$, $\mathbf{p}(I_{f})$, ${\mathbf{p}_{}}_{\mathrm{BH}}(I_{f})$, and $\mathbf{C}(I_{f})=\mathbf{C}(0)$
\end{algorithmic}
\end{algorithm}

\subsection{Particle Swarm Optimization for $\mathcal{PB}$}\label{subsec_optprobsol_pso}
%***************PARAGRAPH MESSAGE START***********************
%Next paragraphs present the proposed PSO algorithm. The algorithm is summarized in Algorithm 2 at the end of this section.
%***************PARAGRAPH MESSAGE END*************************
We inherit the power allocations, user-BS associations and UAV 3D locations from step 24 in Algorithm \ref{Algo_1} and use them as initial settings for the PSO algorithm. Using PSO, we find the 3D hovering locations of UAVs and update power allocations accordingly given fixed user-BS associations. In PSO, the swarm moves along multi-dimensional search space in a probabilistic mechanism to find a feasible set of solutions taking into account the movement velocity of the current iteration and the distance between the current position, the position of the best local objective value and the position of the global best objective value~\cite{PSO1}. 

Now, let us consider the movement velocities of $M$ particles that represent the $n^{\text{th}}$ variable at $i^{\text{th}}$ iteration as $\mathbf{v}_n^{(i)}=\left({v_{}}_{n,m}^{(i)}:{m}\in{\mathscr{M}}\right)$. Then the matrix of velocities of $M$ particles can be denoted by $\mathbf{V}^{(i)}=\left[\mathbf{v}_1^{(i)},\dots,\mathbf{v}_N^{(i)}\right]^\top$, where ${\mathbf{V}^{(i)}}\in{\mathbb{R}^{\left({N}\times{M}\right)}}$ and $N$ represents the numbers of optimization variables. Similarly, the matrices of current positions and positions of best local objects can be given by $\mathbf{X}^{(i)}=\left[\mathbf{x}_{1}^{(i)},\dots,\mathbf{x}_{N}^{(i)}\right]^\top$ and $\mathbf{X}_l^{(i)}=\left[\mathbf{x}_{1,l}^{(i)},\dots,\mathbf{x}_{N,l}^{(i)}\right]^\top$, respectively, where $\mathbf{x}_{n}^{(i)}=\left({x_{}}_{n,m}^{(i)}:{m}\in{\mathscr{M}}\right)$ and $\mathbf{x}_{n,l}^{(i)}=\left({x_{}}_{n,m,l}^{(i)}:{m}\in{\mathscr{M}}\right)$. Hence, the positions of best local objectives of $M$ particles representing the $n^{\text{th}}$ variable can be given by:
\begin{equation}\label{equ_PSOelemin}
\mathbf{x}_{n,l}^{(i) }=\displaystyle \argmin_{r \leq i}\mathbf{\Theta}\left(\mathbf{x}_n^{(r)}\right)\text{,}    
\end{equation}
where the particle's best local objective is defined among previous $r$ iterations. 

Next, let $\mathbf{x}_g^{(i)}=\left(x_{n,g}^{(i)}:{n}\in{\mathscr{N}}\right)$ represent the positions of global objectives of $N$ variables where $\mathbf{x}_g^{(i)}\in\mathbb{R}^{N \times 1}$ and they are given by: 
\begin{equation}\label{equ_PSOglo}
\mathbf{x}_{g}^{(i)}=\displaystyle \argmin_{m \in \mathscr{M}}\Theta\left({x_{}}_{n,m}^{(i)}\right)\text{,}     
\end{equation}
where $\mathbf{x}_{g}^{(i)}$ is the row-minima of $\mathbf{X}^{(i)}$ and $\Theta$ is the weighted fitness function as we will see in (\ref{equ_PSOfit}). Hence, the movement velocity of $\left(i+1\right)^{\text{th}}$ iteration can be updated as:
\begin{equation}\label{equ_PSOVeUpdt}
    \begin{split}
        \mathbf{V}^{(i+1)}&=\alpha\mathbf{V}^{(i)}+\eta_{1}\mathbf{R}_1\odot\left(\mathbf{X}_{l}^{(i)}-\mathbf{X}^{(i)}\right)\\&+\eta_{2}\mathbf{R}_2\odot\left(\mathbf{x}_{g}^{(i)}-\mathbf{X}^{(i)}\right)\text{,}
    \end{split}
\end{equation}
where the inertia is characterized by $\alpha$ and used to adaptively control the exploration of the optimization process. The cognitive and social learning coefficients are represented by $\eta_{1}$ and $\eta_{2}$, respectively. It is worth noting that, the cognitive and social components in (\ref{equ_PSOVeUpdt}) control the exploration and exploitation of the optimization process. Specifically, exploitation is set to the highest level when $\eta_{1}=0$ and exploration is set to the highest level when $\eta_{2}=0$. Finally, $\mathbf{R}_1,\,{\mathbf{R}_2}\in{\mathbb{R}^{\left({N}\times{M}\right)}}$ are uniformly distributed numbers between $\lbrack 0,1\rbrack$ and $\odot$ denotes the Hadamard product. Consequently, the position of each particle in $(i+1)^{\text{th}}$ iteration can be updated based on its position in $i^{\text{th}}$ iteration and the movement velocity of $(i+1)^{\text{th}}$ iteration as:
\begin{equation}\label{equ_PSOXUpdt}
\mathbf{X}^{(i+1)} \gets \mathbf{X}^{(i)}+\mathbf{V}^{(i+1)}\text{.}
\end{equation}

At each iteration we calculate the difference between received and target SINR as $\boldsymbol{\overline{\upgamma}}_u=\boldsymbol{\upgamma}_u-\epsilon_{\mathrm{u}}$. Now, let us consider the set of users receiving SINR at access and direct links lower than $\epsilon_{\mathrm{u}}$ as $\vartheta_{\mathrm{u}}=\left\{\overline{\upgamma}_u:{\overline{\upgamma}_{u}}\in{\mathbb{R}^{-1}}\right\}$ where $\abs{\vartheta_u}$ denotes the cardinality of $\vartheta_{u}$. Similarly, the set of UAVs receiving SINR at backhaul links lower than $\epsilon_{\mathrm{d}}$ is considered as $\vartheta_{\mathrm{BH}}=\left\{\overline{\upgamma}_{\mathrm{BH}}:{\overline{\upgamma}_{\mathrm{BH}}}\in{\mathbb{R}^{-1}}\right\}$, where $\boldsymbol{\overline{\upgamma}}_{\mathrm{BH}}=\boldsymbol{\upgamma}_{\mathrm{BH}}-\epsilon_{\mathrm{d}}$. Hence, a weighted fitness function can be composed of the objective function and nonlinear inequality constraints in (\ref{equ_PA}) and is given by:
\begin{equation}\label{equ_PSOfit}
    \Theta\left(\mathbf{C},\mathbf{p},{\mathbf{p}_{}}_{\mathrm{BH}}\right)=\mathfrak{R}-\left(e_{1}\abs{\vartheta_u}+e_{2}\abs{\vartheta_{\mathrm{BH}}}\right)\text{,}
\end{equation}
where $e_{1}$ and $e_{2}$ denote penalty parameters and are defined based on the target received QoS at users and UAVs, respectively. $\Theta$ is then evaluated at the current position of each particle and compared with the particle's local best fitness and global fitness of the swarm. The values of $\mathbf{X}_{l}^{(i)}$ and $\mathbf{X}_{g}^{(i)}$ are then updated using~(\ref{equ_PSOelemin}) and~(\ref{equ_PSOglo}), respectively. Although PSO is easy to implement, compared with other evolutionary computation techniques (see, e.g. \cite{GAML} and references therein), the computational complexity of swarm optimization increases with the number of optimization variables and constraints. The weighted fitness function in~(\ref{equ_PSOfit}) reduces the computational complexity of the proposed PSO algorithm and solves the non-linear constrained program in~(\ref{equ_PAPB}) independently of the number of optimization constraints in(~\ref{equ_PA_sub1}). 

The time complexity of PSO can be calculated as follows. $T_{\mathrm{comp}}=T_{\mathrm{int}}+(T_{\mathrm{eva}}+T_{\mathrm{upd}})\times{M}$ where, $T_{\mathrm{int}}$, $T_{\mathrm{eva}}$, $T_{\mathrm{upd}}$, $M$ are the computational costs of the initialization, evaluation, velocity and position update of each particle, and the number of particles respectively~\cite{PSOComp}. Given that the number of optimization variables (i.e., dimensionality of the search space) in Algorithm~2 is $N$, hence, $T_{\mathrm{comp}}=N\left(1+3\times{M}\right)$. Consequently, we denote the complexity of Algorithm~2 as $\mathcal{O}\left(N\times{M}\right)$. The proposed algorithm converges to a near-optimal solution when the relative change in the best objective function value over the last $I_{c}$ iterations is less than $\epsilon_{4}$. The proposed PSO algorithm and time complexity of the proposed algorithms are summarized in Algorithm~\ref{Algo_2} and Table~\ref{tab_timcomp}, respectively.

\begin{algorithm}
\caption{Defines 3D locations of UAVs and updates power allocations accordingly given fixed user-BS associations.}\label{Algo_2}
\begin{algorithmic}[1]  
	\STATE\textbf{Inputs:} user positions, $\mathbf{C}(0)$, $U$, $D$, $\mathbf{p}_{s}^{(\mathrm{max})}$, $N$, $M$, $\alpha$, $\eta_{1}$, $\eta_{2}$, $I_m$, $j=1$, $i=1$
	\STATE\textbf{Initialization:} \\
	$\mathbf{w}(i) \gets \mathbf{w}(I_{f}),\,\mathbf{p}(i) \gets \mathbf{p}(I_{f}),\,{\mathbf{p}_{}}_{\mathrm{BH}}(i) \gets {\mathbf{p}_{}}_{\mathrm{BH}}(I_f)$,\\ $\mathbf{C}(i)\gets\mathbf{C}(I_{f}),\,\mathbf{y}= {\left[\vect(\mathbf{C}(i)),\mathbf{p}(i),{\mathbf{p}_{}}_{\mathrm{BH}}(i) \right]}^\top$,\\ $\mathbf{X}^{(i)}\sim\mathcal{U}\lbrack\epsilon_{1}\mathbf{y},\epsilon_{2}\mathbf{y}\rbrack,\,\mathbf{X}_{l}^{(i)}=\displaystyle\argmin_{r\leq i}\mathbf{\Theta}\left(\mathbf{X}^{(r)}\right)$,\\ $\mathbf{x}_{g}^{(i)}=\displaystyle\argmin_{m\in\mathscr{M}}\Theta\left(x_{n,m}^{(i)}\right)$,
	\WHILE{$j,i \leq I_m$}
	\STATE Compute $\mathbf{V}^{(i)}$, $\mathbf{X}^{(i)}$, $\mathbf{X}_{l}^{(i)}$, $\mathbf{x}_{g}^{(i)}$, $\mathbf{\Theta}^{(i)}\left(\mathbf{C},\mathbf{p},{\mathbf{p}_{}}_{\mathrm{BH}}\right)$ 
		\IF {$\mathbf{\Theta}\left(\mathbf{X}^{(i)}\right)< \mathbf{\Theta}\left(\mathbf{X}_{l}^{(i)}\right)$}
		\STATE $\mathbf{X}_{l}^{(i)}\gets\mathbf{X}^{(i)}$
		    \FOR{$n \in \mathscr{N}$}
		    \IF {$\mathbf{\Theta}\left(\mathbf{x}_{n}^{(i)}\right)< \Theta\left(x_{n,g}^{(i)}\right)$}
		    \STATE ${x_{}}_{n,g}^{(i)} \gets {x_{}}_{n,m}^{(i)}$
		    \ENDIF
		    \ENDFOR
		\ENDIF
		\STATE Update $\mathbf{V}^{(i+1)}$ and $\mathbf{X}^{(i+1)}$ using (\ref{equ_PSOVeUpdt}) and (\ref{equ_PSOXUpdt}), respectively.
		\STATE $i \gets i+1$
		\STATE\textbf{Convergence check:}
		\IF {$\frac{\abs{\Theta^{(i)}\left(\mathbf{C},\mathbf{p},{\mathbf{p}_{}}_{\mathrm{BH}}\right)-\Theta^{(i-I_{c}+1)}\left(\mathbf{C},\mathbf{p},{\mathbf{p}_{}}_{\mathrm{BH}}\right)}}{\abs{\Theta^{(i)}\left(\mathbf{C},\mathbf{p},{\mathbf{p}_{}}_{\mathrm{BH}}\right)}}\leq\epsilon_{4}$, ${i}>{I_{c}}$}
		\STATE $j=0$
		\ENDIF
	\ENDWHILE
	\RETURN $\mathbf{C}(I_{P})$, $\mathbf{p}(I_{P})$, ${\mathbf{p}_{}}_{\mathrm{BH}}(I_{P})$, and $\mathbf{w}(I_{P})=\mathbf{w}(I_{f})$
\end{algorithmic}
\end{algorithm}

\begin{table}[H]
    %\makegapedcells
  \centering
  \caption{Time complexity of the proposed algorithms}\label{tab_timcomp}
        \begin{tabular}{|c|c|}
        %\begin{tabular}{|>{\centering}m{3.2cm}|>{\centering}m{2.2cm}|>{\centering\arraybackslash}m{2cm}|}
        \hline
         %\toprule
            Algorithm  & Time complexity\\[2pt]
        \hline
            Fixed-point method & geometric rate with $\norm{\mathbf{p}_{c}(i)-\mathbf{p}_{c}^{*}}_{\infty}<Ck^{i}$\\[2pt]
        \hline
            PSO & $\mathcal{O}\left(N\times{M}\right)$\\[2pt]
        \hline
        %\bottomrule
        %
        \end{tabular}
        \vspace{-.15in}
\end{table}

\subsection{General Solution for $\mathcal{P}$}\label{subsec_optprobsol_GF}
The design parameters of in-band UAV-assisted IAB networks are intertwined together due to the full reuse of wireless channel resources between backhaul and access links, LOS capabilities of UAVs, small inter-site distance and spatial dynamics of user distribution. Hence, we present an iterative algorithm in Algorithm~\ref{Algo_3} that combines Algorithm~\ref{Algo_1} and Algorithm~\ref{Algo_2} to solve problem~(\ref{equ_PA}). Let us consider $(i>1)$ in Algorithm~\ref{Algo_3}. Hence, the proposed algorithm updates user-BS association vector $\mathbf{w}(i)$ based on the 3D locations matrix $\mathbf{C}(i-1)$. Then, the set of different user-BS associations between current and previous iteration is defined in step~\ref{UE-BSchk}. If $\norm{\mathbf{w}(i)-\mathbf{w}(i-1)} \ge \epsilon_{1}$ for some $\epsilon_{1}\ge 0$, then, a new set of 3D locations is obtained in step~\ref{NewLocs}. In other words, a new iteration of Algorithm~\ref{Algo_2} is required for convergence. Similarly, the set of different 3D locations of UAVs and the sum rate difference are obtained in steps~\ref{Locchk} and~\ref{ratechk}, respectively to define whether new iteration of Algorithm~\ref{Algo_1} is required for convergence. The proposed algorithm converges to a near-optimal feasible set of solutions after a few iterations. 

%***************PARAGRAPH MESSAGE START***********************
%How is the proposed algorithm different from previous studies?
%***************PARAGRAPH MESSAGE END*************************
Our proposed solution for UAV-assisted IAB networks is significantly different compared to the studies in~\cite{BHUAV1BS,UAVBHMBSs,UAVIoT, FPIDLMaxMin, PSOAccess}. In particular, we consider the mutual dependence between backhaul, direct and access transmissions, inter-cell interference and the mutual dependence between the spatial configurations of UAVs and the spatial dynamics of ground user distribution, which are significantly challenging in UAV-based cellular scenarios.

\begin{algorithm}
\caption{General fixed-point iteration and PSO algorithm }\label{Algo_3}
\begin{algorithmic}[1]  
	\STATE\textbf{Inputs} $I_m$, $j=1$, $i=1$
	\WHILE{$j,i \leq I_m$}
	\STATE Compute $\mathbf{w}$, $\mathbf{p}$, ${\mathbf{p}_{}}_{\mathrm{BH}}$ using Algorithm (\ref{Algo_1}) 
		\IF {$i\not=1$}
		    \IF {$\norm{\mathbf{w}(i)-\mathbf{w}(i-1)} \le \epsilon_{5}$ for some $\epsilon_{5} \ge 0$} \label{UE-BSchk}
		    \STATE break
		    \ENDIF
		\ENDIF
		\STATE Compute $\mathbf{C}$ and update $\mathbf{p}$ and ${\mathbf{p}_{}}_{\mathrm{BH}}$ accordingly using Algorithm (\ref{Algo_2}) \label{NewLocs}
		\IF{$i\not=1$}
		    \IF {$\norm{\mathbf{C}(i)-\mathbf{C}(i-1)} \le \epsilon_{6}$ for some $\epsilon_{6} \ge 0$} \label{Locchk}
		    \STATE break
		    \ENDIF
		\ELSE
		        \IF {$\left|{\mathfrak{R}(i)-\mathfrak{R}(i-1)}\right| \le \epsilon_{7}$ for some $\epsilon_{7} \ge 0$} \label{ratechk}
		        \STATE break
		        \ENDIF
		\ENDIF
		\STATE $i \gets i+1$
	\ENDWHILE
	\RETURN $\mathbf{C}(I_{G})$, $\mathbf{v}(I_{G})$, $\mathbf{p}(I_{G})$ and ${\mathbf{p}_{}}_{\mathrm{BH}}(I_{G})$
\end{algorithmic}
\end{algorithm}

\section{Drone Antenna Array Spatial Configuration}\label{sec_DAA}
%***************PARAGRAPH MESSAGE START***********************
%Motivation for using the drone antenna array (DAA).
%***************PARAGRAPH MESSAGE END*************************
In previous sections, we presented how a group of UAVs can be spatially configured as distributed IAB-nodes to serve multiple hotspots for in-band IAB scenarios. As the number of ground users increases, the number of required UAVs for coverage enhancement and capacity boosting increases as well, entailing more design challenges and higher levels of interference between direct, access and backhaul links. Moreover, the SINR formulas in~(\ref{equ_SINR_b>d}),~(\ref{equ_SINR_b>t}) and~(\ref{equ_SINR_d>a}) show that decreasing the inter-site distance poses more technical challenges in the design of the proposed in-band IAB drone network architecture. To this end, we consider another spatial configuration mode for UAVs. In that, UAVs are configured as a single DAA to serve ground users that are spatially distributed in a single hotspot. Unlike distributed UAVs, UAVs in DAA mode are not interfering to each other, but are rather composed in a single antenna array to benefit from the potential advantages of the DAA~\cite{DAAJor}. The DAA configuration mode allows for on-demand array configurations. Specifically, the design parameters of the DAA are adjusted based on the spatial distribution of ground users to maximize the overall sum rate gains.

%drone, i.e., antenna, element gains, and array directivity.
% =======
% FIG. 02
% =======
\begin{figure}
  \begin{center}
  \captionsetup{justification=centering}
  \includegraphics[width=7cm,height=7cm,keepaspectratio]{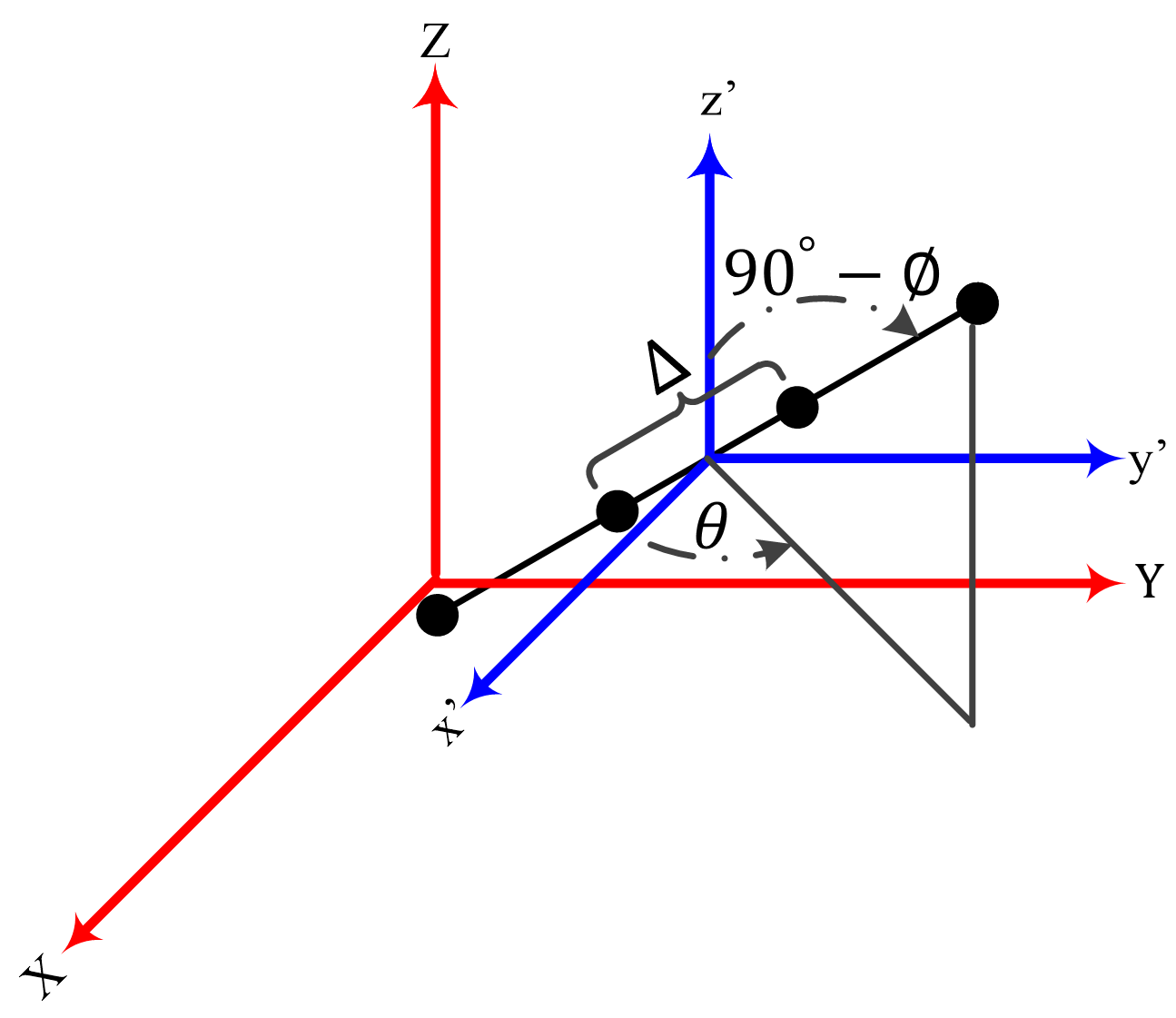}%\vspace*{-5mm} to decrease vertical space between caption and figure
  \caption{Drone antenna array design parameters.}\label{fig_DAA}
  \end{center}
  \vspace{-.25 in}
\end{figure}
\subsection{Backhaul Downlink Transmissions}\label{subsec_sysmodDAA_BHDLTXions}
%***************PARAGRAPH MESSAGE START***********************
%MISO channels and LZFBF MU-MIMO precoder in DAA mode. This paragraph also presents how channel coefficients (and consequently sum rate (objective function) and SINR expressions (optimization constraints)) can be defined in terms of the DAA design parameters.  
%***************PARAGRAPH MESSAGE END*************************
The MISO channel between DAA $\mathrm{r}$, composed of single antenna $D$ drones and $a^{\text{th}}$ aUE ${\mathbf{h}_{\mathrm{r},a}}\in{\mathbb{C}^{\left({1}\times{D}\right)}}$ is given by: 
\begin{equation}\label{equ_DAA_hA>a}
\mathbf{h}_{\mathrm{r},a} = \frac{1}{\sqrt{D}}\times{\left[h_{1,a},\dots,h_{D,a}\right]}\text{,}
\end{equation}
where ${h_{d,a}}\in{\mathbb{C}^{({1}\times{1})}}$ is the access link channel coefficient between $d^{\text{th}}$ antenna, i.e., drone, element and $a^{\text{th}}$ aUE. It follows the same definition as (\ref{equ_hb>t}). Let us consider the set of DAA design parameters as $\mathcal{X}$ where $\mathcal{X}=\left\{\theta,\phi,\Delta_{\mathrm{r}},x_c,y_c,z_c\right\}$. ${\theta}\in{\left[0,2\pi\right]}$, ${\phi}\in{\left[0,2\pi\right]}$, $\Delta_{\mathrm{r}}$ and $\left[x_c,y_c,z_c\right]$ are the azimuth angle from $(x^\prime,z^\prime)$ plane, elevation angle from $(x^\prime,y^\prime)$ plane, antenna element separation and 3D coordinates of the DAA center, i.e., coordinates of the origin of $(x^{\prime},y^{\prime},z^{\prime})$ plane, respectively (see Fig.~\ref{fig_DAA}). Now the 3D coordinates of $d^{\text{th}}$ drone element in the DAA can be defined in terms of $\mathcal{X}$ and they are given by:  
\begin{equation}\label{equ_dist(DAAParams)}
    \begin{split}
        \left[x_d,y_d,z_d\right] &=\left[x_c,y_c,z_c\right]+{\frac{\Delta_{\mathrm{r}}\left(D-2d+1\right)}{2}}\\&\times{\left[\mathrm{cos}(\phi)\mathrm{cos}(\theta),\mathrm{cos}(\phi)\mathrm{sin}(\theta),\mathrm{sin}(\phi)\mathrm{sin}(\theta)\right]}\text{.}
    \end{split} 
\end{equation}

Consequently, $\mathbf{h}_{\mathrm{r},a}$ can be defined in terms of $\mathcal{X}$ as $\mathbf{h}_{\mathrm{r},a}(\mathcal{X})$ and is used to define LZFBF precoder for multi-user MISO transmissions at access links of the DAA. The LZFBF precoder at the DAA is given by ${\mathbf{V}_{\mathrm{r}}}\in{\mathbb{C}^{({D}\times{L})}}=\mathbf{H}_{\mathrm{r}}^\dagger=\mathbf{H}_{\mathrm{r}}^*\left[\mathbf{H}_{\mathrm{r}}^{}\mathbf{H}_{\mathrm{r}}^*\right]^{-1}$, where ${\mathbf{H}_{\mathrm{r}}}\in{\mathbb{C}^{({L}\times{D})}}$ is the full rank channel matrix between DAA and $L$ aUEs with $\mathbf{H}_{\mathrm{r}}(\mathcal{X})=\left[{\mathbf{h}_{}}_{\mathrm{r},1}(\mathcal{X}),\dots,\mathbf{h}_{\mathrm{r},L}(\mathcal{X})\right]^{\top}$. By utilizing the DAA configuration mode, DAA divides aUEs into spatial division multiple access (SDMA) groups. In that, the set of SDMA group of aUEs that are associated with the DAA and scheduled at same time and spectrum resources is represented by $\mathscr{L}$ where $\abs{\mathscr{L}}=L$. 

It is worth noting that, the spatial multiplexing gains are constrained by the number of drones in DAA. In particular, the DAA exploits LZFBF to transmit $L$ independent spatial streams for downlink access transmissions, where $L\leq D$. Now, let us consider DAA that is configured to serve a group of ground users that are spatially distributed away from IAB-donor and concentrated in the center of a single hotspot. Hence, the received signal at $d^\text{th}$ antenna element from IAB-donor can be written as: 
\begin{equation}\label{equ_DAA_b>d}
    \begin{split}
        {y_{}}_{\mathrm{b},d}&=\underbrace{\sqrt{{p_{}}_{\mathrm{b},d}}{\mathbf{h}_{}}_{\mathrm{b},d}{\mathbf{v}_{}}_{\mathrm{b},d}{x_{}}_{\mathrm{b},d}}_{\text{transmitted signal}}+\underbrace{\sum_{j \in \mathscr{D}\backslash d}\sum_{i\in\mathscr{L}}\sqrt{{p_{}}_{\mathrm{r},i}}{\mathbf{h}_{}}_{j,d}{\mathbf{v}_{}}_{\mathrm{r},i}{x_{}}_{\mathrm{r},i}}_{\text{self-interference}}\\&+\underbrace{\sum_{j \in \mathscr{D}\backslash d}\sqrt{{p_{}}_{\mathrm{b},j}}{\mathbf{h}_{}}_{\mathrm{b},d}{\mathbf{v}_{}}_{\mathrm{b},j}{x_{}}_{\mathrm{b},j}}_{\text{inter-stream interference}}+n_d\text{,}
    \end{split}
\end{equation}
where $\mathscr{L}$ is the SDMA group of interfering aUEs to  $t^{\text{th}}$ tUE . The second and third term in~(\ref{equ_DAA_b>d}) denote the self-interference and inter-stream interference on the backhaul transmissions of the DAA. The received SINR at $d^\text{th}$ drone can be defined as:
\begin{equation}
    \label{equ_DAASINR_b>d}
    {\upgamma_{}}_{\mathrm{b},d} =\frac{{p_{}}_{\mathrm{b},d}\abs{{\mathbf{h}_{}}_{\mathrm{b},d}{\mathbf{v}_{}}_{\mathrm{b},d}}^{2}}{\sigma^{2}}\text{.}
\end{equation}

\subsection{Access Downlink Transmissions}\label{subsec_sysmodDAA_AccDLTXions}
Similarly, the received downlink signal and SINR at $t^{\text{th}}$ tUE from IAB-donor are given by:
\begin{equation}\label{equ_DAA_b>tUE}
    {y_{}}_{\mathrm{b},t}=\underbrace{\sqrt{{p_{}}_{\mathrm{b},t}}{\mathbf{h}_{}}_{\mathrm{b},t}{\mathbf{v}_{}}_{\mathrm{b},t}{x_{}}_{\mathrm{b},t}}_{\text{transmitted signal}}+\underbrace{\sum_{i\in\mathscr{L}}\sqrt{{p_{}}_{\mathrm{r},i}}{\mathbf{h}_{}}_{\mathrm{r},t}{\mathbf{v}_{}}_{\mathrm{r},i}{x_{}}_{\mathrm{r},i}}_{\text{inter-tier interference}}+n_t\text{,}
\end{equation}
\begin{equation}\label{equ_DAASINR_b>t}
    {\upgamma_{}}_{\mathrm{b},t}=\frac{p_{b,t}\abs{{\mathbf{h}_{}}_{\mathrm{b},t}{\mathbf{v}_{}}_{\mathrm{b},t}}^{2}}{\displaystyle\sum_{i\in{\mathscr{L}}}{p_{}}_{\mathrm{r},i}{\abs{{\mathbf{h}_{}}_{\mathrm{r},t}{\mathbf{v}_{}}_{\mathrm{r},i}}}^{2}+\sigma^{2}}\text{,}
\end{equation}    
respectively, where $\mathscr{D} \cup \overline{\mathscr{T}}$ denotes the set of interfering direct and backhaul link transmissions to $a^{\text{th}}$ UE and make interference. Finally, the received downlink signal and SINR at $a^{\text{th}}$ aUE from DAA are defined  as~(\ref{equ_DAA_d>aUE}) and~(\ref{equ_DAASINR_D>a}), respectively where: 
\begin{equation}\label{equ_DAA_d>aUE}
    {y_{}}_{\mathrm{r},a}=\underbrace{\sqrt{{p_{}}_{\mathrm{r},a}}{\mathbf{h}_{}}_{\mathrm{r},a}{\mathbf{v}_{}}_{\mathrm{r},a}{x_{}}_{\mathrm{r},a}}_{\text{transmitted signal}}+\underbrace{\displaystyle\sum_{{k}\in{\mathscr{D}\cup\overline{\mathscr{T}}}}\sqrt{{p_{}}_{\mathrm{b},k}}{\mathbf{h}_{}}_{\mathrm{b},a}{\mathbf{v}_{}}_{\mathrm{b},k}{x_{}}_{\mathrm{b},k}}_{\text{inter-tier interference}}+n_a\text{,}
\end{equation}
\begin{equation}\label{equ_DAASINR_D>a}
    {\upgamma_{}}_{\mathrm{r},a} =\frac{{p_{}}_{\mathrm{r},a}\abs{{\mathbf{h}_{}}_{\mathrm{r},a}{\mathbf{v}_{}}_{\mathrm{r},a}}^{2}}{\displaystyle\sum_{k\in\mathscr{D}\cup\overline{\mathscr{T}}}{p_{}}_{\mathrm{b},k}\abs{{\mathbf{h}_{}}_{\mathrm{b},a}{\mathbf{v}_{}}_{\mathrm{b},k}}^{2}+\sigma^{2}}\text{.}
\end{equation}

\subsection{Network Sum Rate Maximization}\label{subsec_ProbDAA}
%***************PARAGRAPH MESSAGE START***********************
%This paragraph presents how the sum rate maximization problem can be defined in terms of the array design parameters. And how it can be solved
%***************PARAGRAPH MESSAGE END*************************
Next, we show how the network performance can be improved in in-band IAB scenarios by spatially configuring UAVs as a single DAA. The network sum rate maximization problem is given by:
\begin{equation}\label{equ_OptDAA}
\max_{\mathcal{X},\mathbf{w},\mathbf{p},{\mathbf{p}_{}}_{\mathrm{BH}}} \quad \mathbf{1}_A^\top\log_2(1+{\boldsymbol{\upgamma}_{}}_\mathscr{A})+\mathbf{1}_T^\top\log_2(1+{\boldsymbol{\upgamma}_{}}_\mathscr{T})\text{,}\\
\end{equation}
\begin{subequations}
    \begin{align}
        \label{equ_DAAsub1}
            \mathrm{subject \, to} \quad & {\boldsymbol{\upgamma}_{}}_\mathscr{U} \ge{\epsilon_{\mathrm{u}}}\text{,\,} {\boldsymbol{\upgamma}_{}}_\mathscr{D}\ge{\epsilon_{\mathrm{d}}}\text{,}\\ 
        \label{equ_DAAsub2}
            &\abs{\Delta_{d+1}^{(\mathrm{r})}-\Delta_{d}^{(\mathrm{r})}}\geq\Delta_{\mathrm{r}}^{(\mathrm{min})}\text{,}\,\forall\,d\in\mathscr{D}\text{,}\\ 
        \label{equ_DAAsub3}
            &\theta \in \lbrack 0,2\pi\lbrack, \phi \in \lbrack 0,2\pi\lbrack, \\
        \label{equ_DAAsub4}
            &\mathbf{m}\le\mathbf{p}_{\mathscr{S}}^{(\mathrm{max})}\text{,} 
    \end{align}
\end{subequations}
where the minimum separation between the DAA antenna elements is defined in~(\ref{equ_DAAsub2}) as $\Delta_{\mathrm{r}}^{(\mathrm{min})}$ to avoid collisions. As shown in~(\ref{equ_OptDAA}), the problem is cast in terms of $\mathcal{X}$ and is independent of the number of antenna elements of the DAA. In DAA-assisted in-band IAB scenarios, the network performance enhancement is directly proportional to the number of antenna elements of the DAA (see Section~\ref{subsec_res_daa}). Hence, it is of paramount importance to design problem~(\ref{equ_OptDAA}) such that its computational complexity is independent of the number of UAVs. Problem~(\ref{equ_OptDAA}) shares the same logarithmic objective function and SINR non-linear inequality constraints as~(\ref{equ_PA}). Hence, it is solved using the two-stage iterative algorithm in Algorithm~\ref{Algo_3}. Finally, $\mathcal{PA}$ and $\mathcal{PB}$ can be defined as~(\ref{equ_PAPA_DAA}) and~(\ref{equ_PAPB_DAA}), respectively where:
\begin{equation}\label{equ_PAPA_DAA}
    \begin{split}
        \mathcal{PA:} &\min_{\mathbf{w}, \mathbf{p}, {\mathbf{p}_{}}_{\mathrm{BH}}} \quad \mathbf{1}_U^\top \mathbf{p}+\mathbf{1}_D^\top {\mathbf{p}_{}}_{\mathrm{BH}}\text{,}\\&\mathrm{subject\,to}~\text{(\ref{equ_DAAsub1}) and (\ref{equ_DAAsub4})}\text{,}
    \end{split}
\end{equation}

%\begin{equation}\label{equ_PAPA_DAA}
%    \begin{split}
%        \mathcal{PA:} &\max_{\mathbf{w}, \mathbf{p}, {\mathbf{p}_{}}_{\mathrm{BH}}} \quad \mathbf{1}_A^\top \log_2\left(1+{\boldsymbol\upgamma_{}}_\mathscr{A}\right)+\mathbf{1}_T^\top \log_2\left(1+{\boldsymbol\upgamma_{}}_\mathscr{T}\right)\text{,}\\&\mathrm{subject\,to}~\text{(\ref{equ_DAAsub1}) and (\ref{equ_DAAsub4})}\text{,}
%    \end{split}
%\end{equation}
\begin{equation}\label{equ_PAPB_DAA}
    \begin{split}
        \mathcal{PB:} &\max_{\mathcal{X}, \mathbf{p}, {\mathbf{p}_{}}_{\mathrm{BH}}} \quad \mathbf{1}_A^\top \log_2\left(1+{\boldsymbol\upgamma_{}}_\mathscr{A}\right)+\mathbf{1}_T^\top \log_2\left(1+{\boldsymbol\upgamma_{}}_\mathscr{T}\right)\text{,}\\&\mathrm{subject\,to}~\text{(\ref{equ_DAAsub1}) - (\ref{equ_DAAsub4})}\text{.}
    \end{split}
\end{equation}

\begin{table}
    %\makegapedcells
  \centering
  \caption{Simulation parameters}\label{tab_simparams}
        \begin{tabular}{|c|c|c|}
        %\begin{tabular}{|>{\centering}m{3.2cm}|>{\centering}m{2.2cm}|>{\centering\arraybackslash}m{2cm}|}
        \hline
         %\toprule
            Settings  & Distributed UAVs & Single DAA\\[2pt]
        \hline
            IAB-donor TM: direct links & SISO & SISO\\[2pt]
        \hline
            IAB-donor: backhaul links & MIMO & MIMO\\[2pt]
        \hline
            IAB-donor: TX antennas & $64$ & $64$\\[2pt]
        \hline
            Number of UAVs & $4$ & $4$\\[2pt]
        \hline
            UAV: TX antennas & $1$ & $1$\\[2pt]
        \hline
            DAA TM: access links & $-$ & MIMO ($4$ layers)\\[2pt]
        \hline
            UAV TM: access link & SISO & $-$\\[2pt]
        \hline
            Number of users & $25$ & $25$\\[2pt]
        \hline
        $f_c$, $\mathrm{BW}$\!, $p_{\mathrm{b}}^{(\mathrm{max})}$\!, $p_{d}^{(\mathrm{max})}$ & \multicolumn{2}{c|}{$2~\mathrm{GHz}$, $20~\mathrm{MHz}$, $46~\mathrm{dBm}$, $36~\mathrm{dBm}$}\\[2pt]
        \hline
            $\sigma^{2}$\!, $\epsilon_{\mathrm{u}}$, $\epsilon_{\mathrm{d}}$ & \multicolumn{2}{c|}{$-104~\mathrm{dBm}$, $3~\mathrm{dB}$, $3~\mathrm{dB}$}\\[2pt]
        \hline
            $M$\!, $\alpha$, ${\eta_{}}_{1}$\!, ${\eta_{}}_{2}$ & \multicolumn{2}{c|}{$200,\,\left[ 0.1,1.1\right]\!,\,1.49,\, 1.9$}\\[2pt]
        \hline
        %\bottomrule
        %
        \end{tabular}
        \vspace{-.15in}
\end{table}

\section{Numerical Results}\label{sec_results}
%***************PARAGRAPH MESSAGE START***********************
%Brief summary about section VI.
%***************PARAGRAPH MESSAGE END*************************
In this section, we numerically evaluate the performance gains of using UAVs as IAB-nodes in in-band IAB networks. Specifically, we use Algorithm~(\ref{Algo_3}) and Monte Carlo simulations to study the achievable gains in received downlink SINR and overall network sum rate. In doing so, we define two use cases for the spatial configurations of UAVs based on the spatial distribution of ground users and compare their performance with the baseline scenario, in which, UAVs are not used. In the baseline scenario, we define the downlink access power allocations as ${p_{}}_{\mathrm{b},u}^{*}=\left(\frac{1}{\lambda}-\frac{N_0}{\abs{{h_{}}_{\mathrm{b},u}}^2}\right)^+$\!, where ${p_{}}_{\mathrm{b},u}^{*}$ is the waterfilling power allocation and $\lambda$ satisfies $\frac{1}{U}\sum_{{u}\in{U}}\left(\frac{1}{\lambda}-\frac{N_0}{\abs{{h_{}}_{\mathrm{b},u}}^2}\right)^+=p_{\mathrm{b}}^{(\mathrm{max})}$. Each UAV is equipped with a single transmit antenna due to the limited volume, weight, and payload of drone IAB-nodes. The channel realizations in~(\ref{equ_hb>t}) and~(\ref{equ_steering}), and the spatial distribution of ground users are randomly updated every Monte Carlo simulation. The simulation parameters of both scenarios are summarized in Table~\ref{tab_simparams}.

% =======
% FIG. 03
% =======
\begin{figure}
  \begin{center}
  \captionsetup{justification=centering}
  \includegraphics[width=7cm,height=7cm,keepaspectratio]{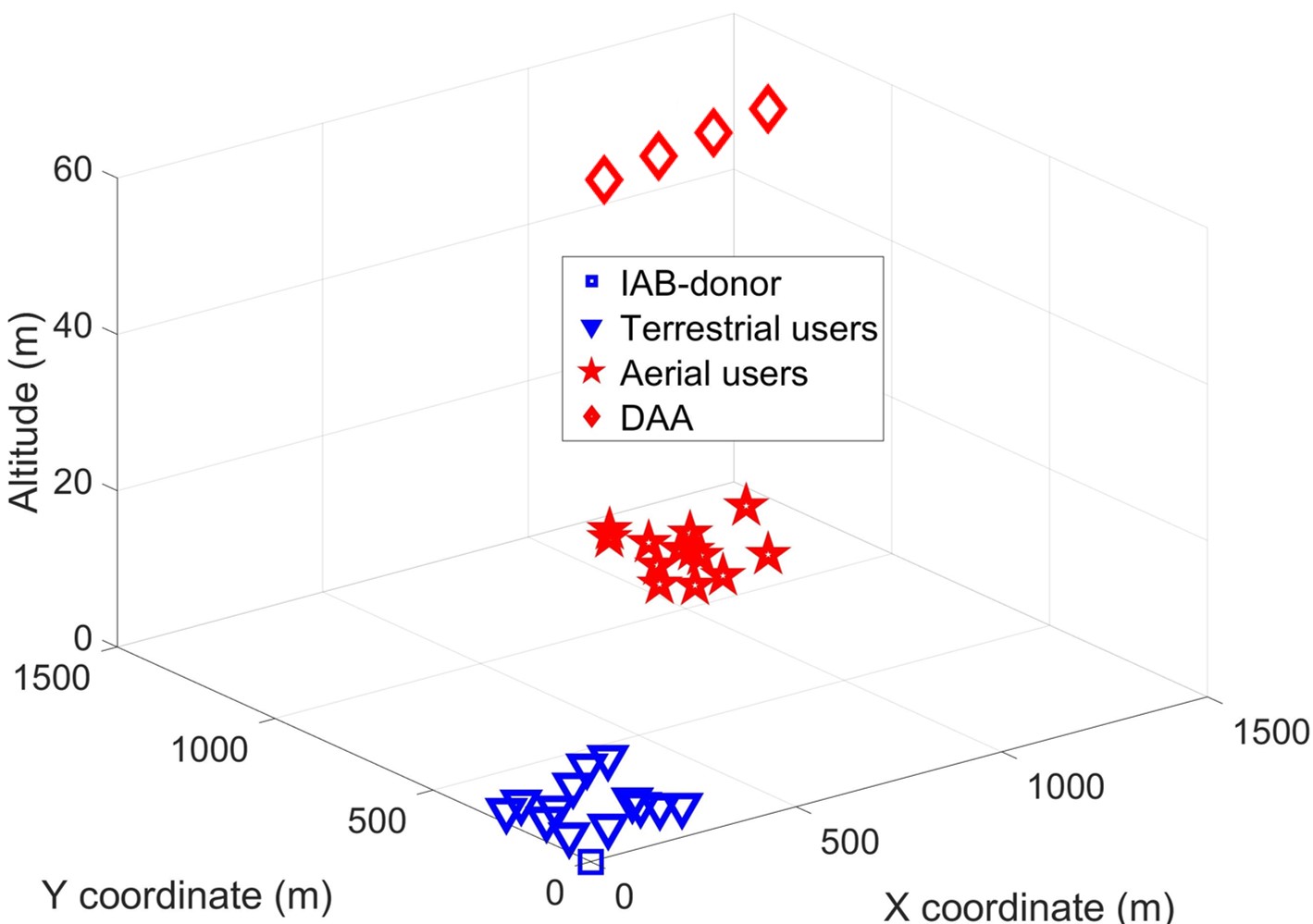}%\vspace*{-5mm} to decrease vertical space between caption and figure
  \caption{Dual clusters: spatial configurations of DAA.}\label{SCMs_UAVs_DAA}
  %\vspace{-.05 in}
  \end{center}
\end{figure}

% =======
% FIG. 04
% =======
\begin{figure}
  \begin{center}
  \captionsetup{justification=centering}
  \includegraphics[width=7cm,height=7cm,keepaspectratio]{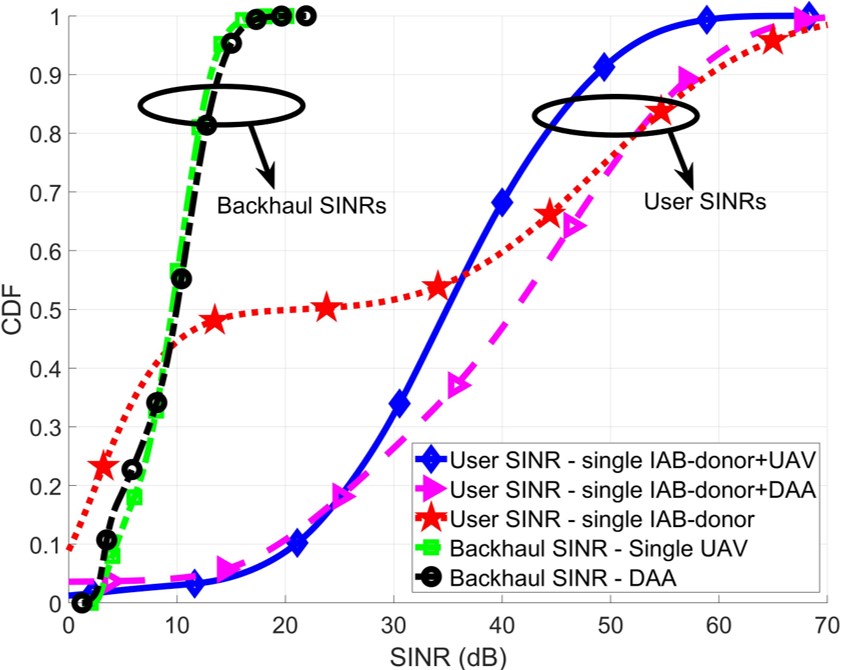}%\vspace*{-5mm} to decrease vertical space between caption and figure
  \caption{Dual clusters: received downlink SINR.}\label{DAA_UeSINR}
  \vspace{-.15in}
  \end{center}
\end{figure}

\subsection{Dual Clusters Spatial Distribution of Cellular Users}\label{subsec_res_daa}
%***************PARAGRAPH MESSAGE START***********************
%Dual clusters: Results of the received downlink SINR/throughput
%***************PARAGRAPH MESSAGE END*************************
In this scenario, we study the use case where users are concentrated in the center of a single hotspot, e.g., music festivals and sports events as depicted in Fig.~\ref{SCMs_UAVs_DAA}. In such scenarios, it is better for aUEs to be associated with a single DAA rather than being associated with distributed UAVs (see Section~\ref{sec_DAA}). Although IAB-donor allows for multi-user MIMO transmissions at backhaul links, it adopts SISO downlink transmissions to tUEs. Hence, we can fairly evaluate the performance of using DAA with different spatial distributions of ground users (see Section~\ref{subsec_res_distuavs}). Fig.~\ref{DAA_UeSINR} shows that the average received SINR of ground users is enhanced by more than $30$ dB after using DAA. Further, it reveals that the received SINR of tUEs is slightly decreased in order to increase the SINR of aUEs. Fig.~\ref{DAA_UeSINR} also shows how the spatial configuration of UAVs is intertwined with the spatial distribution of ground users. In that, the received SINR is significantly improved when UAVs are configured as DAA compared with the spatial configuration of distributed UAVs. Finally, Fig.~\ref{DAA_UeSINR} shows that the received SINR at backhaul links is consistent with the inequality constraints~(\ref{equ_PA_sub1}) and~(\ref{equ_DAAsub1}).  

% =======
% FIG. 05
% =======
\begin{figure}
  \begin{center}
  \captionsetup{justification=centering}
  \includegraphics[width=7cm,height=7cm,keepaspectratio]{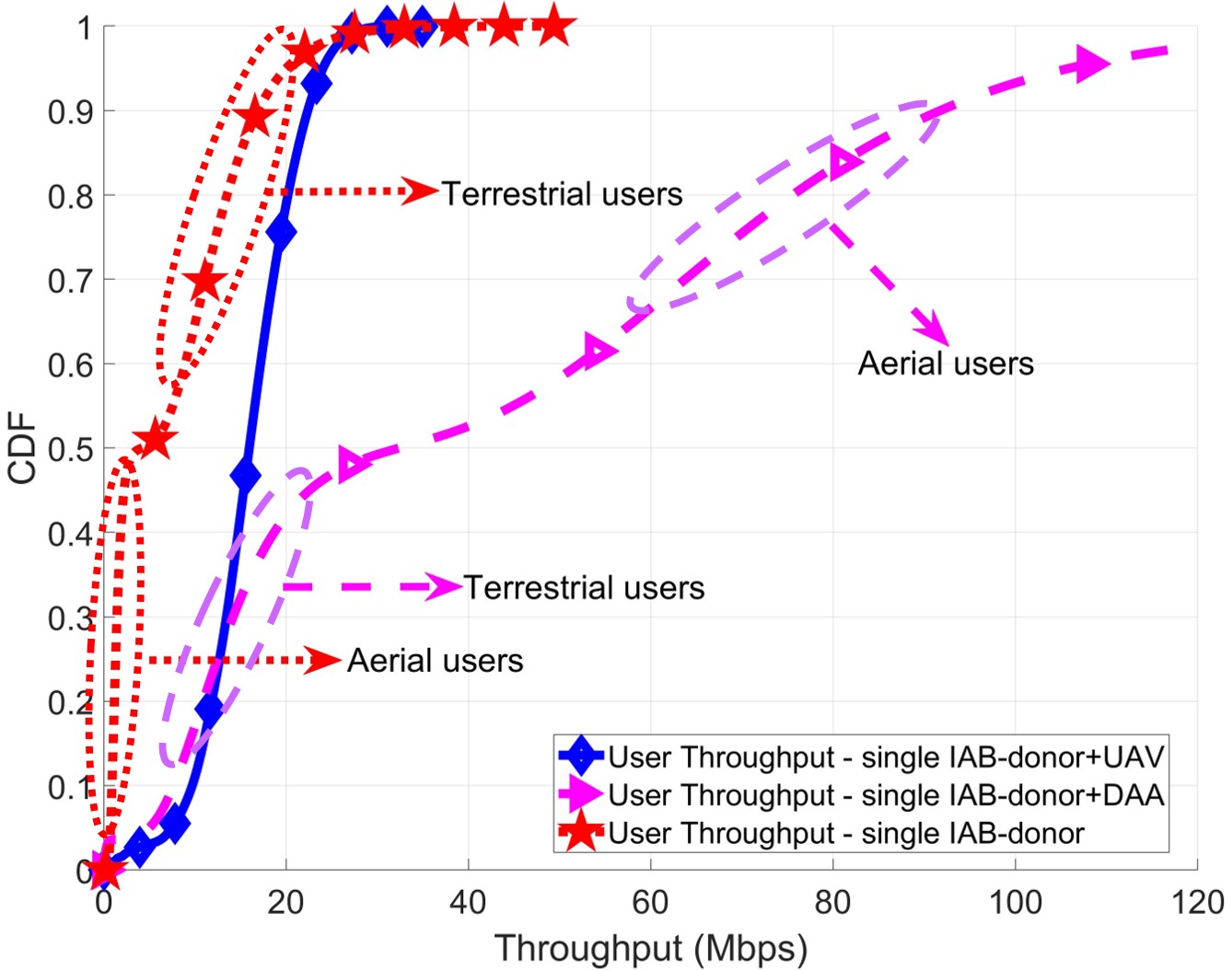}%\vspace*{-5mm} to decrease vertical space between caption and figure
  \caption{Dual clusters: received downlink user throughput.}\label{DAA_UeThr}
  \vspace{-.15in}
  \end{center}
\end{figure}

% =======
% FIG. 06
% =======
\begin{figure}
  \begin{center}
  \captionsetup{justification=centering}
  \includegraphics[width=7.5cm,height=7.5cm,keepaspectratio]{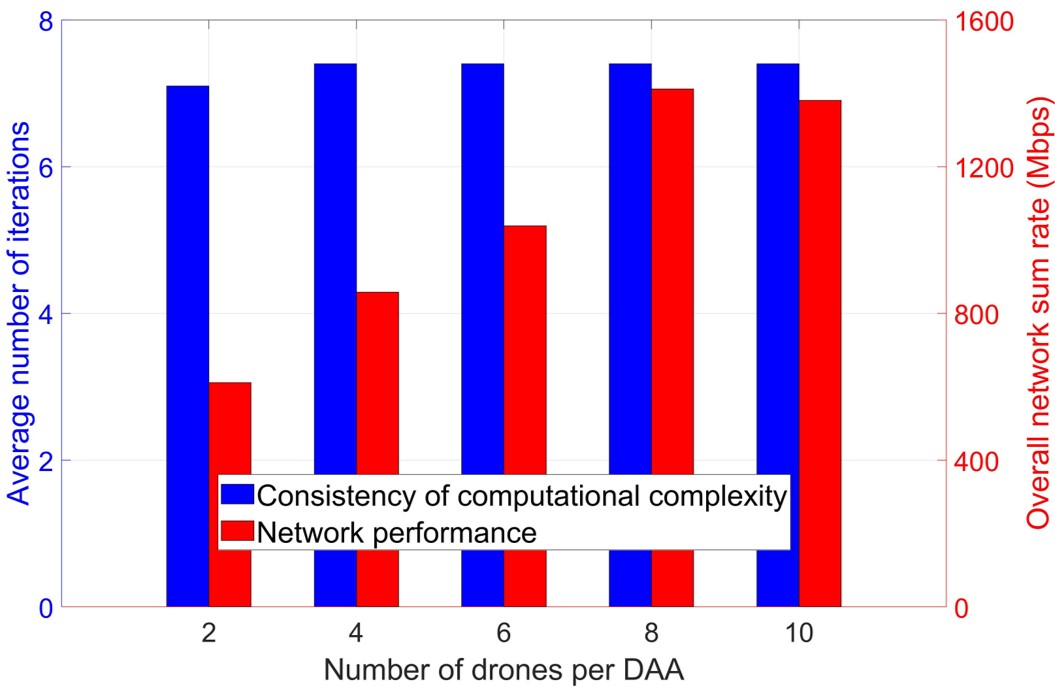}%\vspace*{-5mm} to decrease vertical space between caption and figure
  \caption{The computational complexity of Algorithm~\ref{Algo_2} with respect to the number of drones per DAA.}\label{NumofItrs}
  \vspace{-.15in}
  \end{center}
\end{figure}

The enhancement in the received downlink throughput in Fig.~\ref{DAA_UeThr} is consistent with the results in Fig.~\ref{DAA_UeSINR}. It is worth noting that the received throughput at aUEs is higher than that of tUEs after using the DAA. This is because the use of DAA allows for $D$-fold spatial multiplexing gain. Generally, the DAA exploits full spectrum resources to transmit $D$ independent spatial streams to $D$ users per SDMA group. Hence, the allocated spectrum resources to aUEs are now much higher than those allocated to tUEs. Consequently, Fig.~\ref{DAA_UeThr} reveals that UAVs can be used as DAA in in-band IAB scenarios not only for coverage enhancement but also for capacity boosting. Fig.~\ref{DAA_UeThr} also shows that offloading aUEs from IAB-donor to DAA helps to improve the downlink throughput of tUEs. Finally, it is worth noting that the number of UAVs per DAA can be increased based on the capacity demands, while ensuring the same computational complexity of~(\ref{equ_OptDAA}). 

Fig.~\ref{NumofItrs} demonstrates the consistency of the computational complexity of the proposed algorithm for a larger number of UAVs. The number of iterations is slightly increased due to the increased dimensions of ${\mathbf{p}_{}}_{\mathrm{BH}}$ in~(\ref{equ_PAPB_DAA}). It also shows how the overall network performance is directly proportional to the number of UAVs when they are spatially configured as DAA. Further, it reveals that the network performance decreases at high number of UAVs due to the increased levels of mutual interference between backhaul and access links.

% =======
% FIG. 07
% =======
\begin{figure}
  \begin{center}
  \captionsetup{justification=centering}
  \includegraphics[width=7cm,height=7cm,keepaspectratio]{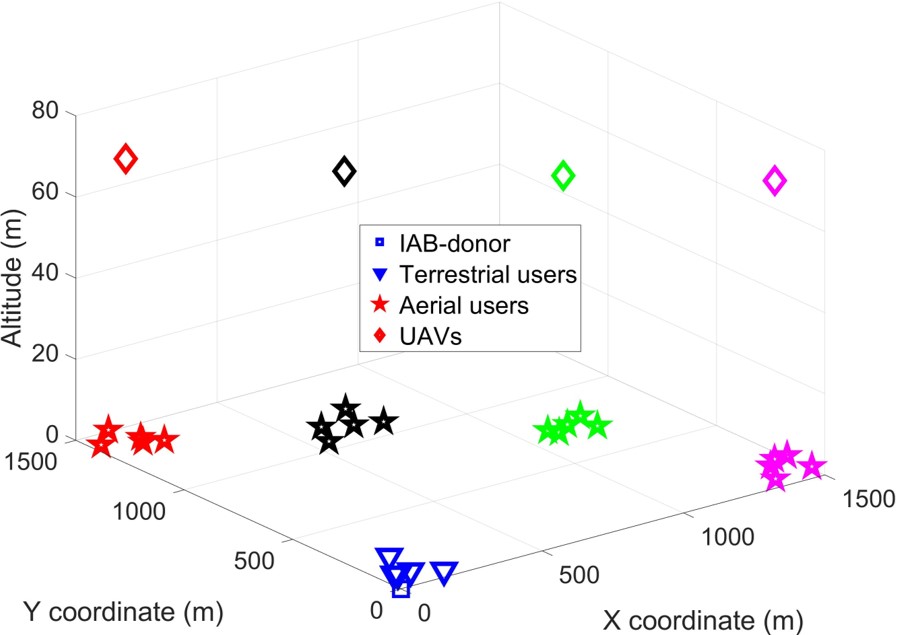}%\vspace*{-5mm} to decrease vertical space between caption and figure
  \caption{Multiple clusters: spatial configurations of UAVs.}\label{SCMs_UAVs_DUAVs}
  %\vspace{-.1in}
  \end{center}
\end{figure}

% =======
% FIG. 08
% =======
\begin{figure}
  \begin{center}
  \captionsetup{justification=centering}
  \includegraphics[width=7cm,height=7cm,keepaspectratio]{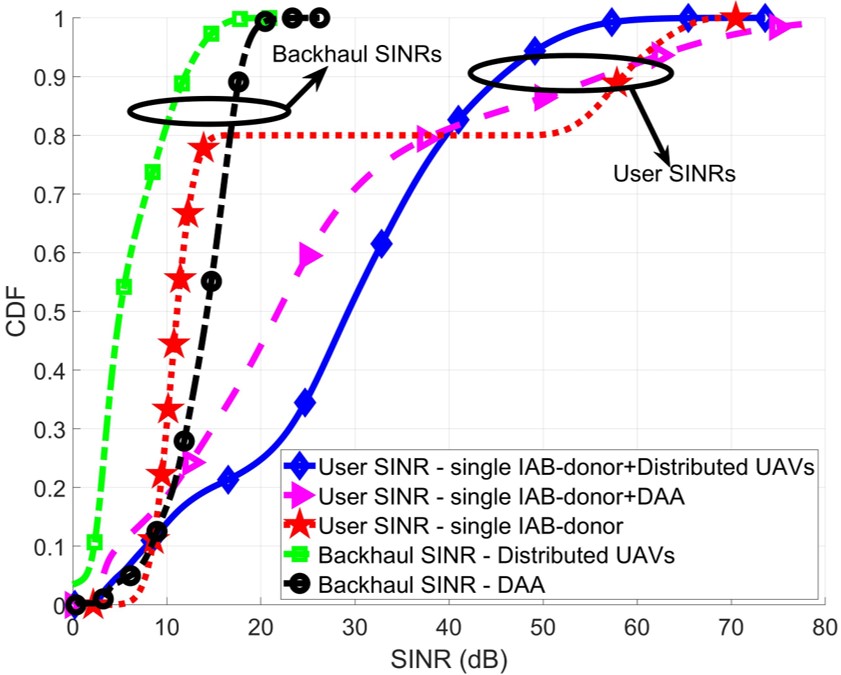}%\vspace*{-5mm} to decrease vertical space between caption and figure
  \caption{Multiple clusters: received downlink SINR.}\label{dist_UeSINR}
  \vspace{-.15in}
  \end{center}
\end{figure}

\subsection{Multiple Clusters Spatial Distribution of Cellular Users}\label{subsec_res_distuavs}
%***************PARAGRAPH MESSAGE START***********************
%multiple clusters: Results of the received downlink SINR/throughput
%***************PARAGRAPH MESSAGE END*************************
In this scenario, users are normally distributed into multiple clusters in the designated coverage area as depicted in Fig.~\ref{SCMs_UAVs_DUAVs}. Fig.~\ref{dist_UeSINR} shows that the received SINR is enhanced after deploying the DAA in an optimized 3D location between the user clusters. Further, it is significantly enhanced by more than $20$ dB when UAVs are used as distributed hovering IAB-nodes. These results are consistent with the results in Fig.~\ref{DAA_UeSINR}, in which, the received SINR at tUEs is slightly decreased in order to increase the received SINR at aUEs. In addition, Fig.~\ref{dist_UeSINR} shows that the received SINR at backhaul links is consistent with the inequality constraints~(\ref{equ_PA_sub1}) and~(\ref{equ_DAAsub1}).  

\begin{figure}
  \begin{center}
  \captionsetup{justification=centering}
  \includegraphics[width=7cm,height=7cm,keepaspectratio]{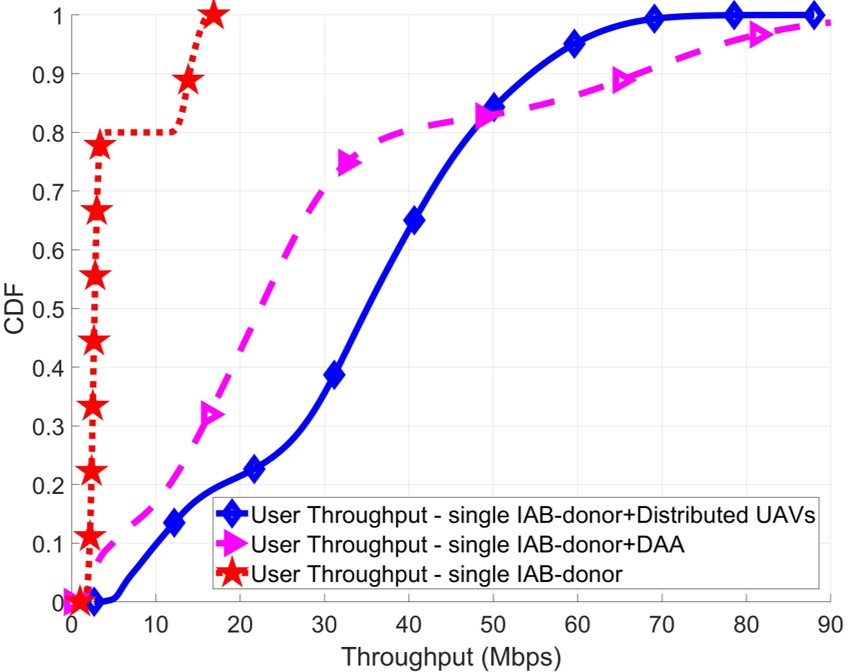}%\vspace*{-5mm} to decrease vertical space between caption and figure
  \caption{Multiple clusters: received downlink user throughput.}\label{dist_UeThr}
  \vspace{-.15in}
  \end{center}
\end{figure}

\begin{figure}
  \begin{center}
  \captionsetup{justification=centering}
  \includegraphics[width=7cm,height=7cm,keepaspectratio]{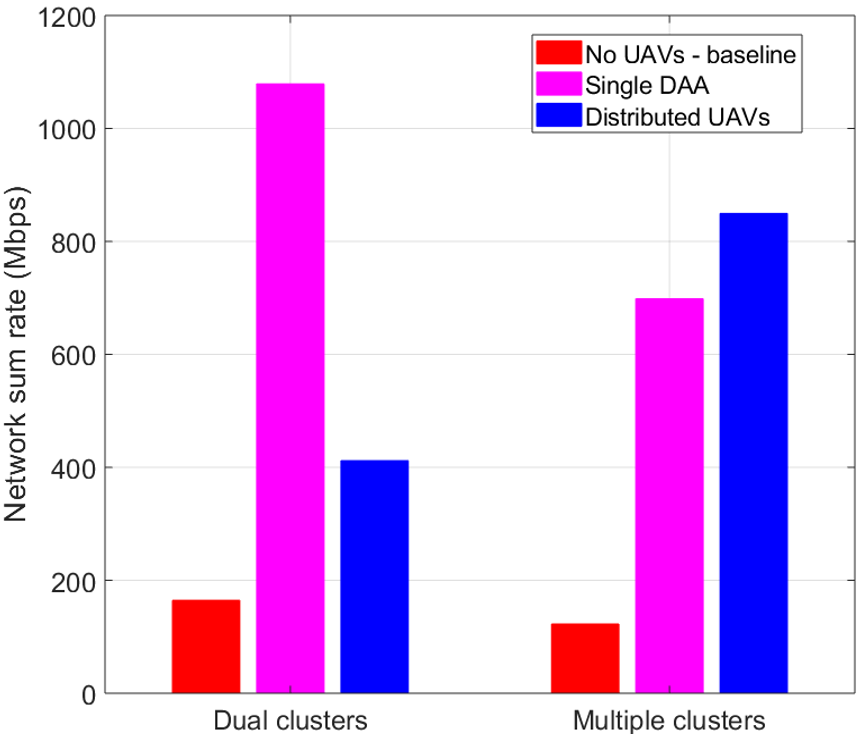}%\vspace*{-5mm} to decrease vertical space between caption and figure
  \caption{Favorable spatial configurations of UAVs.}\label{bars}
  \vspace{-.15in}
  \end{center}
\end{figure}

Fig.~\ref{dist_UeThr} shows that the enhancement in the received downlink throughput is consistent with the results in Fig.~\ref{dist_UeSINR}. It is worth noting that downlink throughput performance of distributed UAVs outperforms that of DAA, although using DAA allows for $D$-fold spatial multiplexing gain. This is because, the low received downlink SINR at aUEs, i.e., users associated with DAA. In particular, the intermediate 3D deployment of DAA between the distributed clusters results in suboptimal directivity towards aUEs. In contrast, the DAA gains are maximized when it is fully directed to serve aUEs concentrated in a single hotspot (as discussed in Section~\ref{subsec_res_daa}). 
Fig.~\ref{bars} presents the favorable spatial configuration of UAVs based on the spatial distribution of ground users.

% =======
% FIG. 09
% =======
% =======
% FIG. 10
% =======

\subsection{Convergence Analysis of the PSO Algorithm}\label{PSOConv}

\begin{figure}
  \begin{center}
  \captionsetup{justification=centering}
  \includegraphics[width=7cm,height=7cm,keepaspectratio]{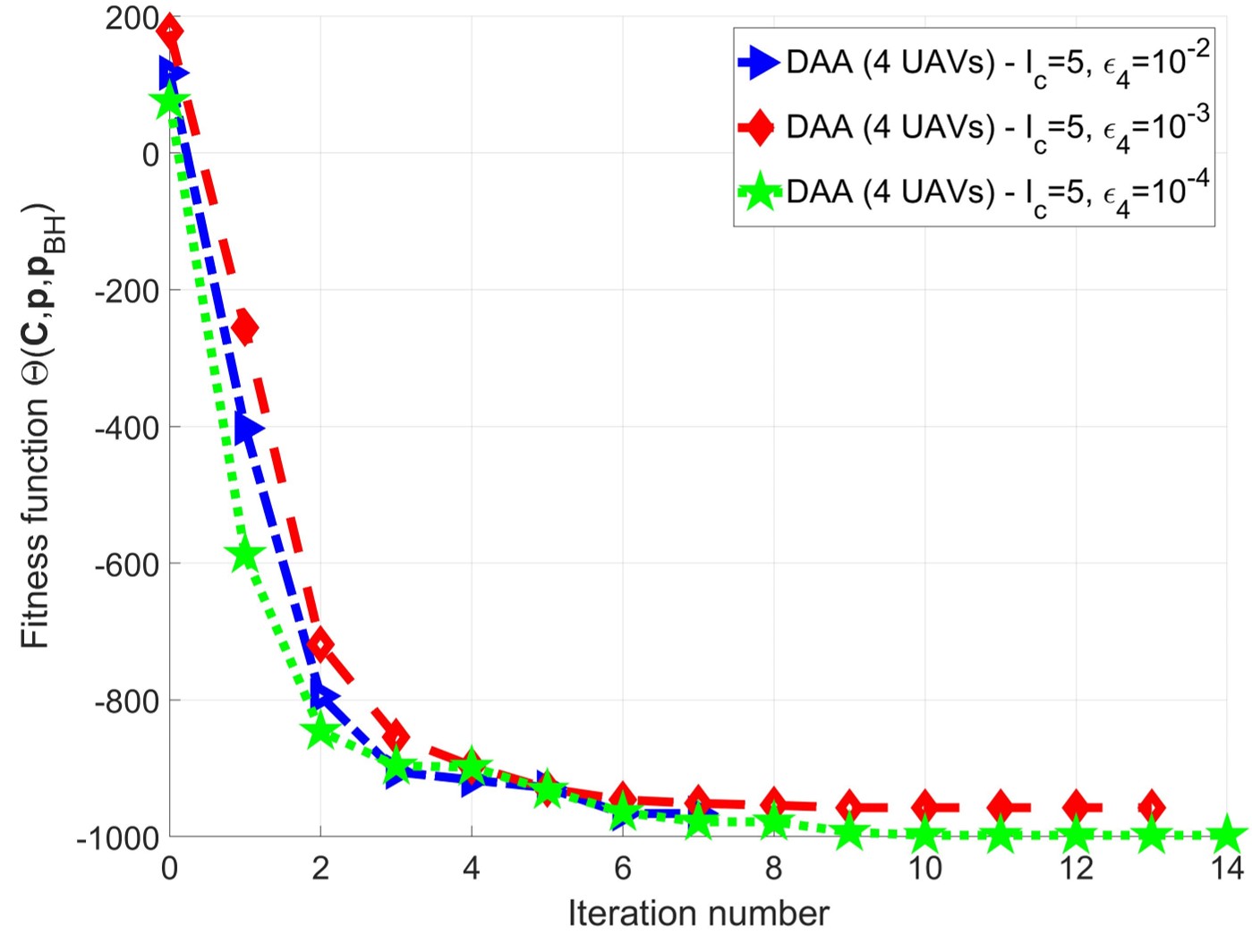}%\vspace*{-5mm} to decrease vertical space between caption and figure
  \caption{Dual clusters: PSO convergence.}\label{PSOCon-DAA}
  \vspace{-.15in}
  \end{center}
\end{figure}

As mentioned in Section~\ref{subsec_optprobsol_pso}, the proposed PSO solution in Algorithm~\ref{Algo_2} converges to a near-optimal solution when the relative change in the best objective function value over the last $I_{c}$ iterations is less than $\epsilon_{4}$. In this section, we analyze the convergence results of the proposed PSO algorithm at different spatial configurations of UAVs. Fig.~\ref{PSOCon-DAA} shows that the fitness function $\Theta\left(\mathbf{C},\mathbf{p},{\mathbf{p}_{}}_{\mathrm{BH}}\right)$ of the proposed PSO algorithm converges to a near-optimal solution after a few number of iterations when UAVs are spatially configured as DAA. It also shows that the time complexity of the proposed PSO algorithm can be significantly improved by increasing the value of $\epsilon_{4}$ without decreasing the accuracy of the optimized set of solutions. 

\begin{figure}
  \begin{center}
  \captionsetup{justification=centering}
  \includegraphics[width=7cm,height=7cm,keepaspectratio]{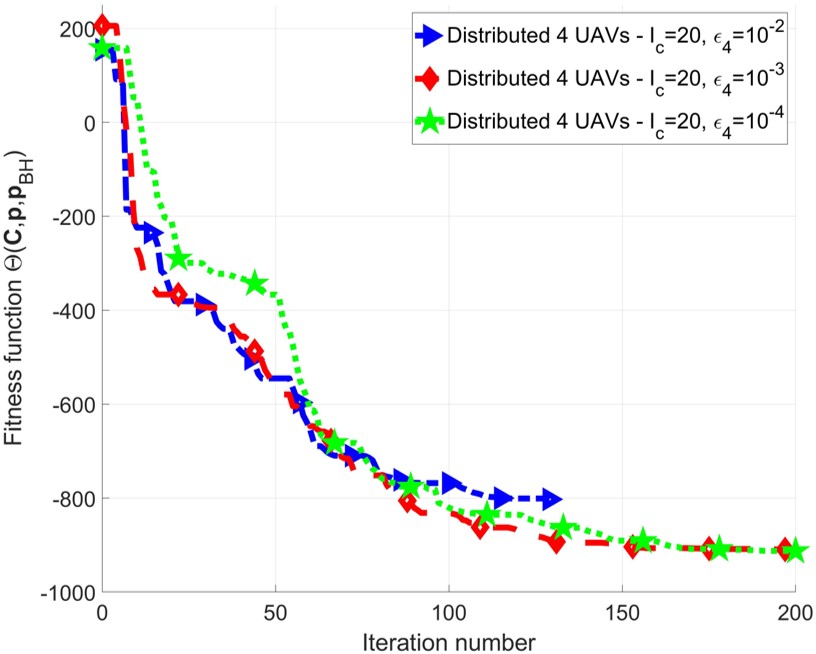}%\vspace*{-5mm} to decrease vertical space between caption and figure
  \caption{Multiple clusters: PSO convergence.}\label{PSOCon-DUAVs}
  \vspace{-.15in}
  \end{center}
\end{figure}

On the other hand, Fig.~\ref{PSOCon-DUAVs} shows that decreasing $\epsilon_{4}$ will impact the accuracy of the optimized set of solutions when UAVs are spatially configured as distributed UAVs (i.e., at a larger number of optimization variables). It is worth noting that the convergence window size (i.e., $I_{c}$) is required to be increased as the number of the optimization variables increases to assure the convergence to a near optimal solution. Hence, we use $I_{c}=5$ and $I_{c}=20$ when UAVs are spatially configured as DAA (Fig.~\ref{PSOCon-DAA}) and as distributed UAVs (Fig.~\ref{PSOCon-DUAVs}), respectively. Finally, Figs.~\ref{PSOCon-DAA} and~\ref{PSOCon-DUAVs} demonstrate that Algorithm~\ref{Algo_2} converges to a near-optimal solution in a fewer number of iterations when UAVs are spatially configured as DAA. In other words, the proposed PSO algorithm converges faster to a near-optimal solution when the number of the optimization variables is smaller. 

% =======
% FIG. 11
% =======

% =======
% FIG. 12
% =======

\subsection{Numerical Evaluation of Reversed Algorithm 3}\label{subsec_res_coupdist}
\begin{figure}
  \begin{center}
  \captionsetup{justification=centering}
  \includegraphics[width=7cm,height=7cm,keepaspectratio]{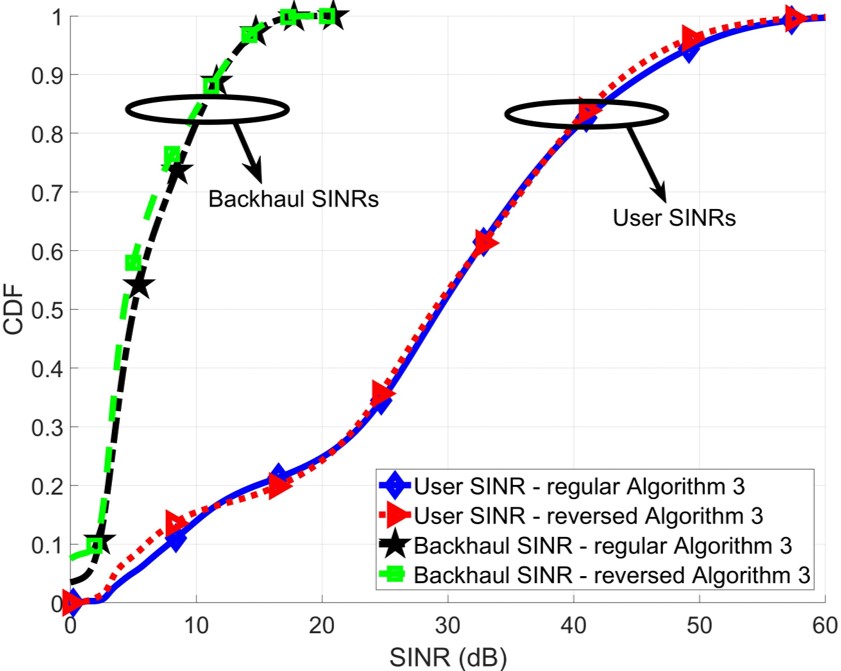}%\vspace*{-5mm} to decrease vertical space between caption and figure
  \caption{Reversed Algorithm~\ref{Algo_3}: downlink SINR.}\label{RvrsdSINR}
  \vspace{-.15in}
  \end{center}
\end{figure}

\begin{figure}
  \begin{center}
  \captionsetup{justification=centering}
  \includegraphics[width=7cm,height=7cm,keepaspectratio]{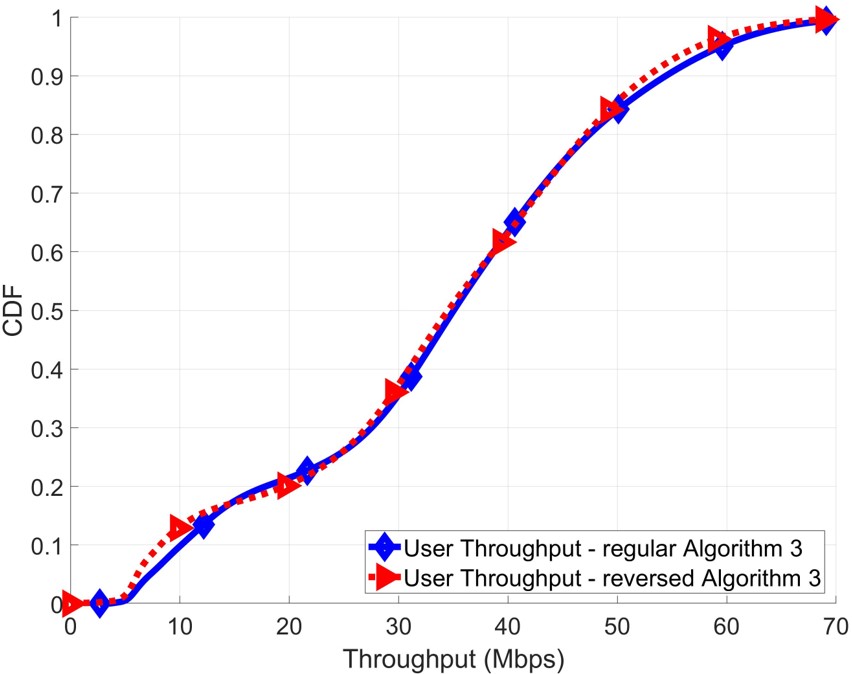}%\vspace*{-5mm} to decrease vertical space between caption and figure
  \caption{Reversed Algorithm~\ref{Algo_3}: downlink throughput.}\label{RvrsdThghpt}
  \vspace{-.22 in}
  \end{center}
\end{figure}

In Section~\ref{subsec_optprobsol_GF}, we presented an iterative solution in Algorithm~3 that combines Algorithms~\ref{Algo_1} and~\ref{Algo_2} to solve the master optimization problem~(9). In this section, we present the numerical results of the reversed version of Algorithm~\ref{Algo_3} (i.e., to optimize the 3D locations of UAVs at first and the user-BS associations at second). We carried out the optimization steps in a reversed order to find the optimized set of solutions when the cellular users are spatially distributed into multiple clusters (see Fig.~7). Our numerical results in Figs.~\ref{RvrsdSINR}~and~\ref{RvrsdThghpt} show that the reversed and regular optimization orders converge to almost the same results. Essentially, the optimized solution of~(9) does not depend on the order of the optimization steps, given that the proposed Algorithm~\ref{Algo_3} converges to a near-optimal set of solutions after a few iterations. However, it is worth noting that the time complexity of the reversed optimization order is always higher than that of the regular order. This is because the PSO algorithm (Algorithm~\ref{Algo_2}) is more time-consuming than the fixed-point method (Algorithm~\ref{Algo_1}). Generally, the number of required PSO iterations in the reversed optimization order is higher than that of the regular order.

% =======
% FIG. 13
% =======

% =======
% FIG. 14
% =======

\subsection{Generic Spatial Distribution of Cellular Users}\label{subsec_res_coupdist}
\begin{figure}
  \begin{center}
  \captionsetup{justification=centering}
  \includegraphics[width=7cm,height=7cm,keepaspectratio]{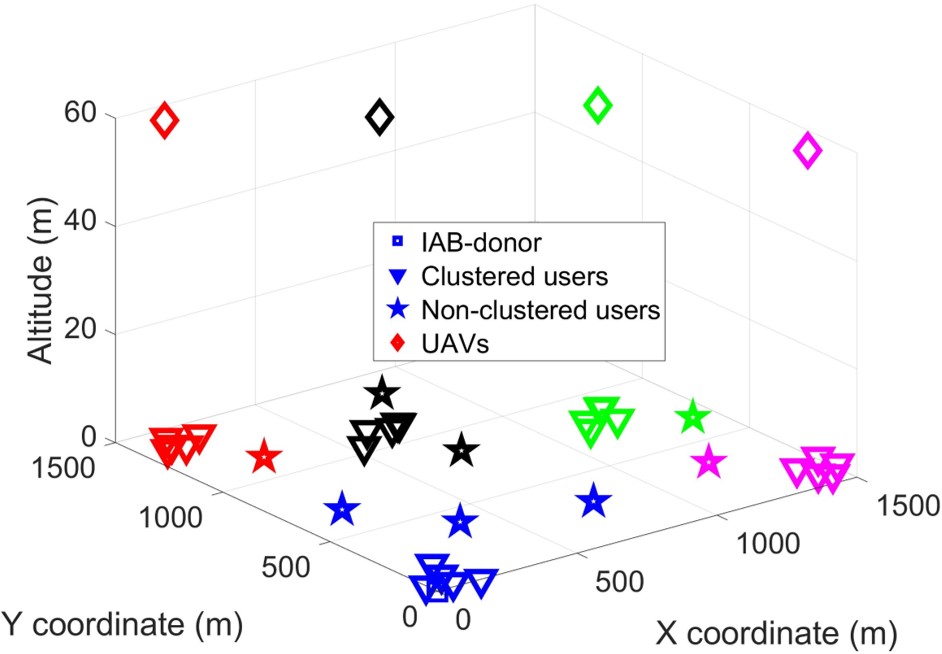}.%\vspace*{-5mm} to decrease vertical space between caption and figure
  \caption{Generic spatial distribution of cellular users.}\label{SCMs_CD}
  %\vspace{-.1in}
  \end{center}
\end{figure}

\begin{figure}
  \begin{center}
  \captionsetup{justification=centering}
  \includegraphics[width=7cm,height=7cm,keepaspectratio]{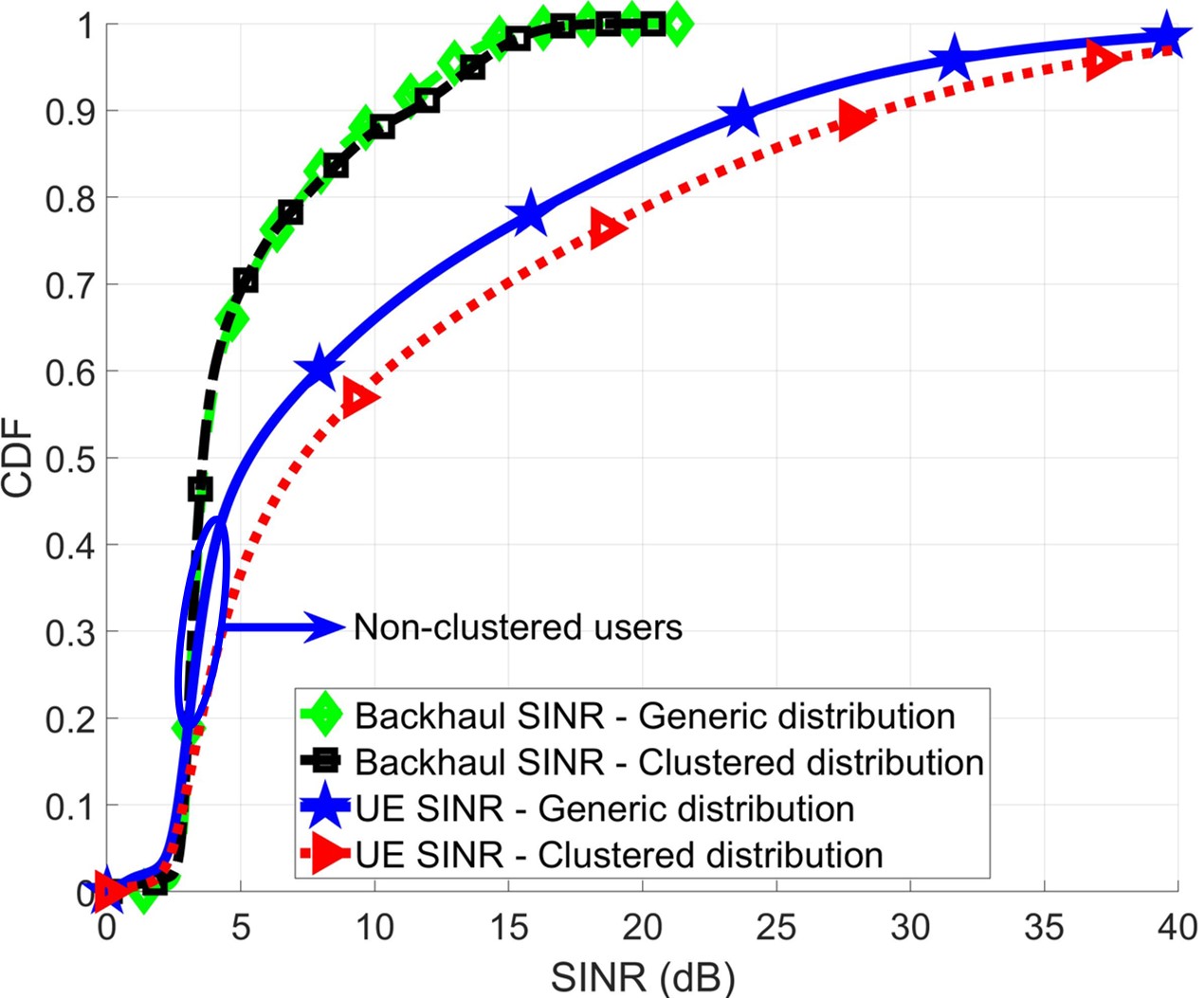}.%\vspace*{-5mm} to decrease vertical space between caption and figure
  \caption{Generic distribution: downlink SINR.}\label{CD_SINR}
  \vspace{-.15in}
  \end{center}
\end{figure}

In this section, we numerically evaluate the performance of generic spatial distribution of cellular users. Specifically, a fraction of users are uniformly distributed within the coverage area (i.e., non-clustered users) and others are distributed into multiple hotspots (i.e., clustered users) as depicted in Fig.~\ref{SCMs_CD}. Fig.~\ref{CD_SINR} shows that the overall received downlink SINR is slightly decreased when cellular users are spatially distributed as clustered and non-clustered users compared with the clustered distribution scenario. Essentially, the received downlink interference levels at non-clustered users are higher than those received at clustered users due to their intermediate locations between the hotspots.

\begin{figure}[t]
  \begin{center}
  \captionsetup{justification=centering}
   \includegraphics[width=7cm,height=7cm,keepaspectratio]{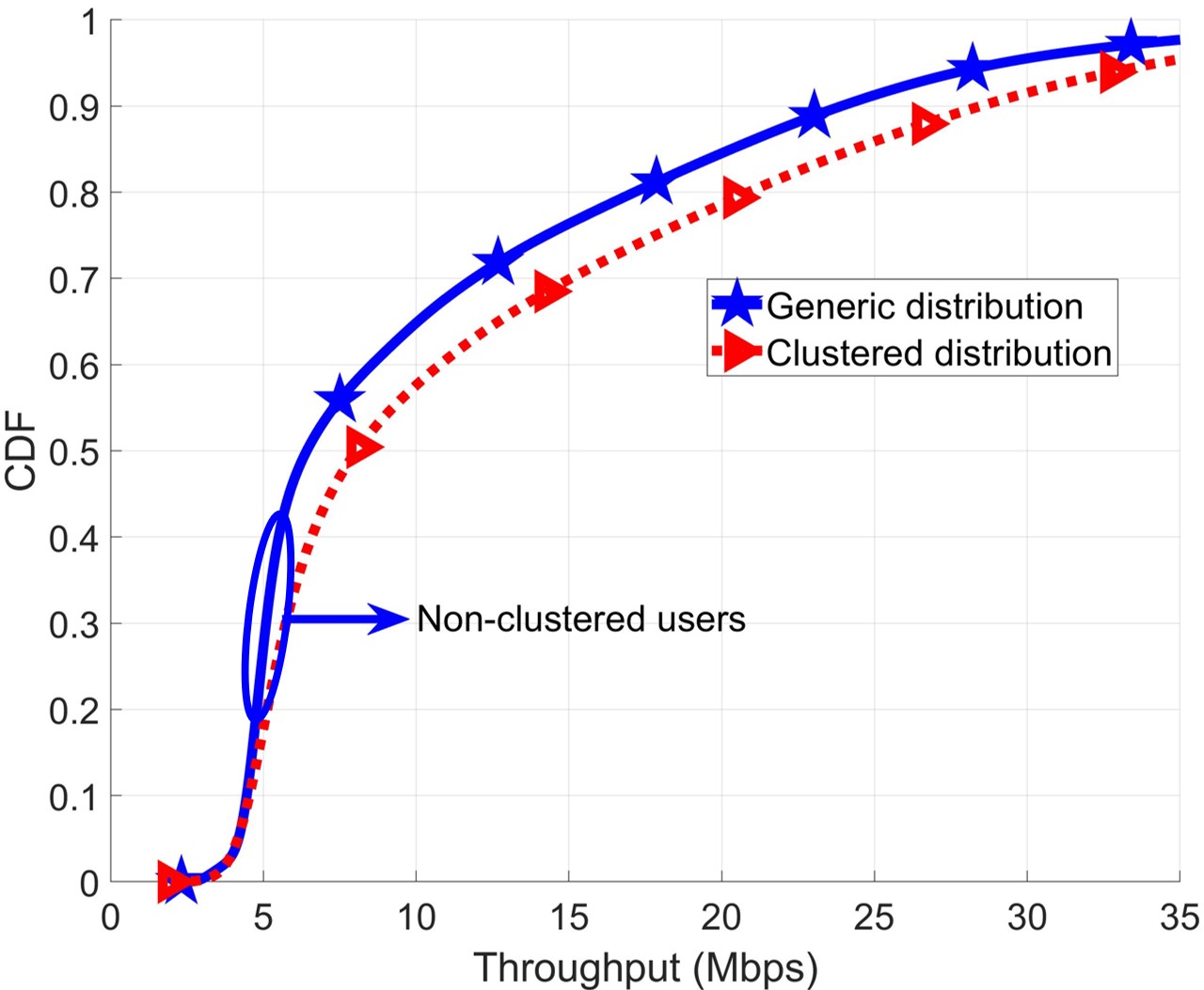}.
  \caption{Coupled distribution: downlink backhaul SINR.}\label{CD_UEThghpt}
  \vspace{-.15in}
  \end{center}
\end{figure}

Thus, the overall SINR performance is decreased by $\approx 2~\mathrm{dB}$ as shown in Fig,~\ref{CD_SINR}. It is worth noting that backhaul performance is almost the same in both scenarios. This is because the spatial distributions of UAVs are almost the same (i.e., the 3D deployment of UAVs). Fig.~\ref{CD_UEThghpt} shows that the downlink throughput is also decreased when the cellular users are spatially distributed into clustered and non-clustered users, which is consistent with the SINR degradation in Fig.~\ref{CD_SINR}. Our numerical results in this section reveal that the performance of the proposed algorithms is directly proportional to the heterogeneity of the spatial distribution of cellular users (i.e., performance gain increases with more clustered users).

% =======
% FIG. 15
% =======

% =======
% FIG. 16
% =======

% =======
% FIG. 17
% =======

% =======
% FIG. 18
% =======
%\begin{figure}
%  \begin{center}
%  \captionsetup{justification=centering}
%  \includegraphics[width=7cm,height=7cm,keepaspectratio]{Figs/33vs25-UEThghpt-BFont2.jpg}.%\vspace*{-5mm} to decrease vertical space between caption and figure
%  \caption{Coupled distribution: downlink user throughput.}\label{CD_UEThghpt}
%  \vspace{-.25in}
%  \end{center}
%\end{figure}

\section{Concluding Remarks}\label{sec_conc}
In this paper, we propose an UAV-based interference management algorithm to optimize the performance of in-band UAV-assisted IAB networks. In-band IAB network architecture allows for tighter interworking between access and backhaul links, making it a promising solution to meet the requirements of fast and easily scalable deployment of next-generation cellular networks. The problem is cast as network sum rate maximization problem. In which, we exploit fixed-point method and PSO to jointly optimize user-BS associations, downlink power allocations and the 3D spatial configurations of UAVs, taking into account the full reuse of wireless channel resources between backhaul, direct and access links, inter-cell interference and LOS capabilities of UAVs. Further, we investigate the mutual dependence between the spatial configurations of UAVs in the sky and the spatial dynamics of ground user distribution. In particular, we consider distributed UAVs and DAA as different spatial configurations of UAVs. 

Our numerical results show that the spatial configuration of distributed UAVs outperforms that of the DAA by $21.6\%$ in terms of the overall network sum rate when the ground cellular users are normally distributed into multiple bad-coverage areas. On the other hand, the spatial configuration of the DAA outperforms that of distributed UAVs by $161.9\%$ when the ground cellular users are concentrated in a single bad-coverage area. Moreover, we show that the proposed algorithm is of low complexity and independent of the number of UAVs when they are spatially configured as DAA. We also analyze the convergence results of the proposed PSO algorithm and show how PSO settings can be adjusted to converge to the same near-optimal set of solutions in fewer number of iterations. We discuss the robustness of the proposed iterative algorithm against the order of the optimization steps and show that it converges to same optimized set of solutions irrespective of the order of the optimization steps. Furthermore, our numerical results reveal that the performance of the proposed algorithms is directly proportional to the heterogeneity of the spatial distribution of cellular users (i.e.,performance gain increases with more clustered users). 

\balance
\bibliographystyle{IEEEtran}
\bibliography{main2.bib,main.bib}

\end{document}